\documentclass[%
 preprint, 
 amsmath,amssymb,
 aps, physrev,
]{revtex4-2}

\usepackage{graphicx}
\usepackage{dcolumn}
\usepackage{bm}
\usepackage{slashed}

\begin{document}


\title{\textbf{Relativistic hydrodynamics of a charged fluid from Myers–Pospelov Lorentz-Violating Electrodynamics.} 
}%

\author{Matheus Duarte}
\email{Contact author: matheus\_duarte@usp.br}
\author{Vitor de Souza}%
\affiliation{%
 Sao Carlos Institute of Physics, University of Sao Paulo, IFSC – USP,
13566-590, Sao Carlos, SP, Brazil.
}%


\date{\today}

\begin{abstract}
We derive the equations of relativistic hydrodynamics for a charged fluid from Lorentz-violating Myers--Pospelov electrodynamics. Starting from the modified fermionic and electromagnetic equations of motion, we employ the covariant Wigner formalism to establish the semiclassical spectral constraints and distribution function, which, combined with Noether's theorem, yield the energy-momentum tensor, particle current, and corresponding conservation laws in the presence of Lorentz violation. The resulting hydrodynamic description contains Lorentz-violating corrections to the thermodynamic quantities and to the fluid currents. In the purely timelike sector, these corrections can be absorbed into effective thermodynamic quantities, preserving the isotropic structure of the fluid equations. In contrast, spacelike backgrounds introduce explicit contributions along the preferred direction, which cannot be absorbed into the standard hydrodynamic variables and give rise to anisotropic forces and torques. These effects lead to departures from standard relativistic charged-fluid hydrodynamics and provide a direct macroscopic manifestation of the Lorentz-violating background. Our results establish a first-principles framework for investigating the macroscopic consequences of Lorentz violation in relativistic charged fluids and provide a starting point for the development of a consistent Lorentz-violating relativistic magnetohydrodynamic theory.
\end{abstract}

\maketitle


\section{Introduction}
\label{sec:intro}

Lorentz invariance \cite{Einstein:1905ve} is a fundamental pillar of modern physics, being central to both the Standard Model of particle physics \cite{ParticleDataGroup:2026aaa} and general relativity \cite{Einstein:1916vd}. It has been extensively tested across a wide range of energies and length scales, providing one of the most successful descriptions of nature to date \cite{Kostelecky:2008ts,Liberati:2013xla}.

Nevertheless, several approaches to quantum gravity, high-energy effective field theories, and theories beyond the Standard Model suggest that Lorentz invariance may not be an exact symmetry at all scales \cite{AlvesBatista:2023wqm,Addazi_2021,Doplicher_1995,Greenberg_2002,Mattingly_2005}. In this context, Lorentz invariance violation (LIV) provides a phenomenological framework to parametrize possible higher-dimensional Lorentz-violating effective operators, suppressed by the Planck scale $\left(E_{\rm Pl} \sim 10^{28} \ \rm{eV} \right)$ \cite{Mattingly_2005}. Such suppressions make terrestrial detection extremely challenging, motivating the use of high-energy astrophysical observations as probes of tiny departures from exact Lorentz symmetry.

Astrophysical messengers such as cosmic rays, gamma rays, and neutrinos are thought to be produced in extreme environments, including active galactic nuclei (AGNs), gamma-ray bursts (GRBs), and relativistic jets \cite{Guepin:2022qpl,Meszaros:2019xej}. The interpretation of these observations requires understanding the plasma environments in which particle acceleration and transport occur. In many of these systems, the plasma dynamics is described by relativistic magnetohydrodynamics (RMHD), which governs the coupled evolution of relativistic fluids and electromagnetic fields \cite{Anile:1989,Hernandez:2017mch}.

Standard RMHD is formulated under Lorentz invariance; therefore, even small violations may induce nontrivial modifications in the dynamics of relativistic plasmas and, consequently, in the evolution of astrophysical sources. Despite this connection, most studies of Lorentz invariance violation in the context of high-energy astrophysics have focused on propagation effects \cite{Amelino-Camelia:1997ieq,PierreAuger:2021tog,Guedes_Lang_2018} and extensive air showers \cite{Guedes_Lang_2018,Saveliev_2023,Duenkel_2021}. More recently, attention has turned to the possible consequences of LIV for particle acceleration and emission processes \cite{Duarte_2024,Duarte_2026,Aguiar:2026eyc}. In parallel, the implications of Lorentz violation for statistical mechanics and equilibrium thermodynamics have also been investigated \cite{Colladay:2004qt,Potting:2022fyw} However, a self-consistent derivation of RMHD equations incorporating LIV at the level of the underlying kinetic theory remains largely unexplored. An important step toward such a formulation is to establish how LIV modify the relativistic hydrodynamics of charged fluids coupled to electromagnetic fields.

In this study, we present a first-principles derivation of the relativistic hydrodynamic equations for a charged fluid in the presence of Lorentz-violating operators in the QED sector. Adopting the Myers--Pospelov framework \cite{Myers:2003fd}, we employ the covariant Wigner formalism \cite{PhysRev.40.749,VASAK1987462} to establish the semiclassical spectral constraints and distribution function. Combined with Noether's conservation laws for the modified energy-momentum tensor and particle current, we derive the corresponding macroscopic hydrodynamic equations and analyze their limiting regimes. Our results establish a theoretical framework for investigating the macroscopic consequences of Lorentz violation in relativistic charged fluids and constitute a first step toward a self-consistent formulation of Lorentz-violating RMHD. Such an extension will require a multicomponent plasma description, including an investigation of the conditions under which conventional RMHD approximations remain applicable in the presence of LIV \cite{Koide:2009yx}.

The paper is organized as follows. In Sec.~\ref{sec:conv}, we present the conventions adopted throughout this work. In Sec.~\ref{sec:liv}, we introduce the Lorentz-violating framework considered. In Sec.~\ref{sec:eom}, we derive the corresponding equations of motion. Section~\ref{sec:wigner} presents the covariant Wigner formalism and its modification in the presence of LIV terms. In Sec.~\ref{sec:eqsRMHD}, we derive the modified relativistic hydrodynamic equations for a charged fluid, while Sec.~\ref{sec:cases} discusses limiting cases for the background vector. In Sec.~\ref{sec:estimates}, we give order-of-magnitude estimates and briefly discuss phenomenology outlooks. Finally, Sec.~\ref{sec:conc} summarizes our conclusions.

\section{Conventions}
\label{sec:conv}

Throughout this work, we use the metric signature
\begin{equation}
g^{\mu\nu} = \mathrm{diag}(1,-1,-1,-1),
\end{equation}
and natural units, $\hbar=c=1$. The Levi-Civita tensor is defined by
$\epsilon^{0123}=+1$, and the dual electromagnetic field tensor is
defined as
\begin{equation}
\tilde{F}^{\mu\nu} =
\frac{1}{2}\epsilon^{\mu\nu\alpha\beta}F_{\alpha\beta}.
\end{equation}
The covariant derivative is defined as
$D_\mu = \partial_\mu + ieA_\mu$.
Repeated Lorentz indices are implicitly summed.

\section{Lorentz violation framework}
\label{sec:liv}

To investigate possible consequences of LIV, we adopt the Myers--Pospelov effective field theory~\cite{Myers:2003fd}. In this framework, dimension-five operators are introduced into the QED Lagrangian, suppressed by the Planck scale $M_{\rm Pl}$.

The resulting effective Lagrangian is
\begin{equation}
\mathcal{L} = \mathcal{L}_{\rm QED}
+ \mathcal{L}^{\rm LIV}_{\gamma}
+ \mathcal{L}^{\rm LIV}_{f},
\end{equation}
where the QED sector is written in its manifestly Hermitian-symmetric Dirac form,
\begin{equation}
\mathcal{L}_{\rm QED} = -\frac{1}{4}F_{\mu\nu}F^{\mu\nu} + \frac{i}{2} \bar{\psi} \gamma^\mu \overleftrightarrow{D}_\mu \psi - m \bar{\psi}\psi,
\end{equation}
with $\bar{\psi}\overleftrightarrow{D}_\mu\psi \equiv \bar{\psi}(D_\mu\psi) - (\overleftarrow{D}_\mu\bar{\psi})\psi$.

The LIV photon-sector contribution is given by
\begin{equation}
\mathcal{L}^{\rm LIV}_{\gamma}
= \frac{\xi}{M_{\rm Pl}}
\, n^{\alpha} F_{\alpha \theta}
\, (n \cdot \partial)
\left(n_{\beta} \tilde{F}^{\beta \theta}\right),
\end{equation}
and the fermionic LIV sector reads
\begin{equation}
\mathcal{L}^{\rm LIV}_{f}
=
\frac{1}{2M_{\rm Pl}}
\, \bar{\psi} \left(\eta_1 \slashed{n} + \eta_2 \slashed{n}\gamma^5 \right) (n \cdot D)^2 \psi
+ {\rm h.c.},
\label{eq:LIV_fermion}
\end{equation}
where $+\mathrm{h.c.}$ denotes the Hermitian conjugate, which is explicitly included in the action and consistently accounted for in the derivation of the equations of motion and the Wigner transport equations. This treatment ensures the Hermiticity of the fermionic action.

The Lorentz-violating background is described by a fixed four-vector $n^\mu$, which selects a preferred spacetime direction. The coefficients $\xi$, $\eta_1$, and $\eta_2$ control the strength of Lorentz violation in the electromagnetic and fermionic sectors, with $\eta_2$ responsible for parity-violating effects.

The Myers--Pospelov framework preserves $U(1)$ gauge invariance despite the presence of Lorentz-violating operators.

Dimension-five Lorentz-violating operators may also raise a naturalness problem, since radiative corrections can induce lower-dimensional operators whose effects are not suppressed by the Planck scale~\cite{Collins:2004bp}. Supersymmetric extensions can mitigate this problem in specific constructions~\cite{GrootNibbelink:2004za,Bolokhov:2005cj}, although the effectiveness of the protection depends on the underlying theory and the supersymmetry-breaking scale~\cite{Redigolo:2011bv}. In the present work, we treat the Myers--Pospelov operators as given effective interactions and focus on their implications for kinetic theory and the hydrodynamics of a relativistic charged fluid. The ultraviolet origin and radiative stability of these operators lie beyond the scope of the present analysis.

\section{Equations of motion}
\label{sec:eom}

As discussed in the previous section, the Myers-Pospelov model introduces higher-derivative operators into the Lagrangian. Consequently, the equations of motion must be derived using the generalized Euler-Lagrange equation, which accounts for derivatives of the fields beyond first order. In the present case, the equations of motion are obtained from

\begin{equation}
\frac{\partial \mathcal{L}}{\partial \phi_i}
- \partial_\mu \left(
\frac{\partial \mathcal{L}}
{\partial (\partial_\mu \phi_i)}
\right)
+ \partial_\mu \partial_\nu
\left(
\frac{\partial \mathcal{L}}
{\partial (\partial_\mu \partial_\nu \phi_i)}
\right)
=0,
\label{eq:euler-lagrange}
\end{equation}

where \(\phi_i\) collectively denotes the dynamical fields of the theory.

\subsection{Electromagnetic sector}

The electromagnetic field satisfies the homogeneous Maxwell equations, expressed by the Bianchi identity,
\begin{equation}
    \partial_\mu \widetilde{F}^{\mu\nu}=0.
    \label{eq:bianchi}
\end{equation}
This identity remains unmodified in the presence of LIV and will be used throughout this work, particularly in the derivation of the hydrodynamic equations.

The equations of motion for the electromagnetic field are obtained by applying Eq.~\eqref{eq:euler-lagrange} to the gauge field $A_\mu$. The resulting modified Maxwell equation reads
\begin{equation}
\partial_\mu F^{\mu\nu}
=
j_{\rm eff}^\nu + \frac{2\xi}{M_{\rm Pl}} \left(n \cdot \partial \right)^2 \left(n_\theta \tilde{F}^{\theta \nu} \right),
\label{eq:maxwell}
\end{equation}
where the effective current is defined as
\begin{equation}
\begin{aligned}
j_{\rm eff}^\nu
&=
j^\nu + j_{\rm LIV}^\nu \\
&=
j^\nu
-
\frac{ie}{M_{\rm Pl}}
n^\nu \bar{\psi}
\left(
\eta_1 \slashed{n}
+\eta_2 \slashed{n}\gamma^5
\right)
\left(
n \cdot \overleftrightarrow{D}
\right)
\psi .
\end{aligned}
\end{equation}
As will be shown in Sec.~\ref{sub:sec:u1}, the effective current naturally arises from the underlying $U(1)$ symmetry and plays a central role in the derivation of the conservation laws.

As originally shown in Ref.~\cite{Myers:2003fd}, the modified Maxwell equation leads to helicity-dependent photon dispersion relations. In the free-field limit, working in the preferred frame $n^\mu=(1,0,0,0)$ and considering photons propagating along the $z$ axis, with four-momentum $k^\mu=(E,0,0,p)$, the transverse circular polarization modes $\epsilon_x\pm i\epsilon_y$ satisfy
\begin{equation}
    \left(E^2-p^2\pm 2\xi\frac{p^3}{M_{\rm Pl}}\right)
    (\epsilon_x\pm i\epsilon_y)=0.
\end{equation}
Here, $E\simeq p$ has been used at leading order in the Planck-suppressed correction. The opposite signs correspond to the two circular polarization states, revealing the birefringent nature of photon propagation in Myers--Pospelov electrodynamics.

\subsection{Fermionic sector}

Applying Eq.~\eqref{eq:euler-lagrange} to the fermionic fields yields the modified Dirac equations. For the spinor field \(\psi\), one obtains
\begin{equation}
\left(
i\slashed{D}
-
m
\right)\psi
=
-\frac{1}{M_{\rm Pl}}
\left(
\eta_1 \slashed{n}
+
\eta_2 \slashed{n}\gamma^5
\right)
\left(
n\cdot D
\right)^2
\psi ,
\label{eq:psi}
\end{equation}
whereas the adjoint spinor satisfies
\begin{equation}
\bar{\psi}
\left(
i\overleftarrow{\slashed D}
+
m
\right)
=
\frac{1}{M_{\rm Pl}}
\bar{\psi}
\left(
\eta_1 \slashed{n}
+
\eta_2 \slashed{n}\gamma^5
\right)
\left(
n\cdot\overleftarrow D
\right)^2 .
\label{eq:psibar}
\end{equation}

As discussed in Ref.~\cite{Myers:2003fd}, the modified Dirac equation leads, in the free-particle limit, to the fermionic dispersion relation
\begin{equation}
    \left[
    E^2-p^2-m^2
    -\frac{2p^3}{M_{\rm Pl}}
    \left(\eta_1+\eta_2\gamma^5\right)
    \right]\psi=0,
    \label{eq:fermion_dispersion}
\end{equation}
where $E\simeq p$ is assumed at high energies. The dispersion relation is helicity dependent, with the two helicity states governed by the effective couplings $\eta_{R,L}=\eta_1\pm\eta_2$ in the ultrarelativistic limit.

Equation~\eqref{eq:psi} constitutes the starting point for the derivation of the modified Wigner equation presented in Sec.~\ref{sec:wigner}. The resulting kinetic description will subsequently be used to obtain the hydrodynamic equations and conservation laws in the presence of Lorentz violation.

\section{Covariant Wigner formalism}
\label{sec:wigner}

Quantum electrodynamics, both in its conventional form and in the presence of Lorentz-violating operators, provides a microscopic description in terms of quantum fields. In order to establish a connection with a kinetic description in phase space, we employ the covariant Wigner formalism, whose central object is the gauge-invariant Wigner function \(W(x,p)\) \cite{PhysRev.40.749,VASAK1987462,Hidaka:2022dmn}.

The construction starts from the gauge-invariant Wigner operator, defined as
\begin{equation}
\hat{W}_{\alpha\beta}(x,p)
= 
\int \frac{d^4y}{(2\pi)^4}e^{-ip\cdot y} 
\bar\psi_{\beta}(x_+)U(A,x_+,x_-)\psi_{\alpha}(x_-),
\end{equation}
where $x_+ \equiv x +\tfrac{1}{2}y$, $x_- \equiv x -\tfrac{1}{2}y$, and $U(A,x_+,x_-)$ is the Wilson line connecting the two spacetime points, ensuring local $U(1)$ gauge invariance \cite{Elze:1987yb,VASAK1987462}. In the Abelian case considered here, it is given by
\begin{equation}
PU(A,x_+,x_-)
=
Pe^{-iey^\mu \int_0^1 ds A_\mu(x_- +sy)} ,
\end{equation}
here $P$ denotes the path ordering along the integration path parameterized by $s$.

The Wigner function is then defined as the ensemble expectation value of the corresponding operator,
\begin{equation}
W(x,p)
=
\left\langle
:\hat{W}(x,p):
\right\rangle,
\end{equation}
where the colons denote normal ordering and the angular brackets represent the ensemble average. The ensemble average is taken with respect to the density operator describing the macroscopic state of the system. Since \(W(x,p)\) is a matrix in Dirac space, the spinor indices will be omitted in the following for notational simplicity.

The covariant Wigner formalism provides a first-principles bridge between the microscopic dynamics encoded in the QED Lagrangian and the macroscopic kinetic and hydrodynamic descriptions \cite{VASAK1987462,Hidaka:2022dmn}. In particular, the moments of the Wigner function determine quantities such as the conserved current and the energy-momentum tensor, which are essential for deriving the macroscopic hydrodynamic equations of a relativistic charged fluid. In the following subsections, we derive the modifications induced by Lorentz-violating operators and obtain the corresponding kinetic equations that constitute the basis of the subsequent hydrodynamic formulation.

\subsection{Equation of motion for the Wigner function}
In order to describe the evolution of the particle ensemble in phase space, we first derive the equation of motion for the Wigner function. Following the procedure of Ref.~\cite{VASAK1987462}, we Fourier transform the quantity
\(i\gamma \cdot (\tfrac{1}{2}\partial_x-\partial_y)\bar{\psi}U\psi\)
and use the modified Dirac equation~\eqref{eq:psi}, obtaining
\begin{equation}
\begin{aligned}
\left [m - \gamma\cdot (p + \tfrac{i\partial_x}{2}) \right]\hat{W}(x,p) =\\
ie \frac{\partial}{\partial p^\mu}\int \frac{d^4y}{(2\pi)^4}e^{-ip\cdot y} \bar \psi (x_+)P\left[\int_0^1 ds \left(1-s\right) F_{\mu \nu}(x_- -sy) U(A,x_+,x_-) \right]\gamma^\nu \psi(x_-) \ + \\
+ \frac{\mathcal{H}}{M_{\rm Pl}}\int \frac{d^4y}{(2\pi)^4}\bar\psi(x_+)PU(A,x_+,x_-)\left(n \cdot D_{x_-} \right)^2 \psi(x_-),
\end{aligned}
\label{eq:W_EOM_1}
\end{equation}
where $\mathcal{H} = (\eta_1 \slashed{n} + \eta_2 \slashed{n}\gamma^5)$. The last term represents the contribution from the Lorentz-violating operators. After repeatedly exploiting the properties of the Wilson line, integrating by parts, and rearranging the resulting covariant derivatives (presented in Appendix~\ref{sec:app1}), Eq.~\eqref{eq:W_EOM_1} can be rewritten as
\begin{equation}
\begin{aligned}
\left[m - \gamma \cdot (p + \tfrac{i\partial_x}{2})+ \frac{\mathcal{H}}{M_{\rm Pl}} \left(n \cdot p + \frac{i n\cdot \partial_x}{2} \right)^2 \right] \hat{W}(x,p) = \\
 ie \left[ \gamma^\nu \frac{\partial}{\partial p^\mu} - \frac{2\mathcal{H}}{M_{\rm Pl}} \left( n \cdot p + \frac{i n\cdot \partial_x}{2} \right) n^\nu \frac{\partial}{\partial p^\mu}\right] \\ \times \int \frac{d^4y}{(2\pi)^4}e^{-ip\cdot y} \bar \psi (x_+)P\left[\int_0^1 ds \left(1-s\right) F_{\mu \nu}(x_- -sy) U(A,x_+,x_-) \right]\gamma^\nu \psi(x_-) \ + \\
 -e\frac{\mathcal{H}}{M_{\rm Pl}}\frac{\partial}{\partial p^\mu}\int \frac{d^4y}{(2\pi)^4}e^{-ip\cdot y} \bar \psi (x_+) P \left[\int_0^1 ds \left(1-s\right)^2 \left(n \cdot \partial \right)n^\nu F_{\mu \nu}(x_- -sy) U(A,x_+,x_-) \right] \psi(x_-) \ + \\
 +e^2 \frac{\mathcal{H}}{M_{\rm Pl}} \left( \frac{\partial}{\partial p^\mu}\right)^2 \int \frac{d^4y}{(2\pi)^4}e^{-ip\cdot y} \bar \psi (x_+)P\left[\int_0^1 ds \left(1-s\right)n^\nu F_{\mu \nu}(x_- -sy) U(A,x_+,x_-) \right]^2 \psi(x_-),
\end{aligned}
\label{eq:W_EOM_2}
\end{equation}
where the gauge-invariant form of the equation becomes explicit.

Following Ref.~\cite{VASAK1987462}, we introduce the operators $\nabla^\mu = \partial_x^\mu - ej_0(\tfrac{\triangle}{2})F^{\mu \nu}\partial_\nu^p$ and $\Pi^\mu = p^\mu - \frac{1}{2}ej_1(\tfrac{\triangle}{2})F^{\mu \nu}\partial_\nu^p$. Employing the $c$-number mean-field (Hartree) approximation, a natural baseline for isolating collective dynamics in ideal, collisionless plasmas, we arrive at the compact form of the equation of motion for the Wigner function:
\begin{equation}
\begin{aligned}
\left[\gamma \cdot \left( \Pi + \tfrac{i\nabla}{2} \right) - m \right]W(x,p) = \frac{\mathcal{H}}{M_{\rm Pl}}\left[n \cdot \left(\Pi + \tfrac{i\nabla}{2} \right) \right]^2 W(x,p) \ + \\
+\frac{\mathcal{H}}{M_{\rm Pl}}e\left\{ \frac{i}{6}j_0 (\tfrac{\triangle}{2}) + \frac{1}{2}j_1 (\tfrac{\triangle}{2}) - \frac{i}{3}j_2 (\tfrac{\triangle}{2}) \right\}\left[\tfrac{i n\cdot \partial_x}{2}; n_\mu F^{\mu \nu} \right] \partial_\nu^p \ W(x,p),
\end{aligned}
\label{eq:W_EOM_3}
\end{equation}
where $j_i(z)$ are the spherical Bessel functions, $\triangle = \partial^p \cdot \partial_x$ and $[a;b]$ denotes the commutator of $a,b$. The intermediate algebra leading from Eq.~\eqref{eq:W_EOM_2} to Eq.~\eqref{eq:W_EOM_3} is presented in Appendix~\ref{sec:app1}.

Equation~\eqref{eq:W_EOM_3} reveals two qualitatively distinct LIV contributions. The first modifies the kinetic operator governing the propagation of the Wigner function through the addition of the term proportional to $n \cdot \left( \Pi + \frac{i\nabla}{2} \right)$, thereby introducing Planck-suppressed corrections to the underlying dispersion relation. The second arises from a commutator involving derivatives and the electromagnetic field tensor. This contribution is of higher order in \(\hbar\) than the terms retained in the transport equation and therefore does not affect the transport dynamics at the order considered here.

\subsection{Clifford decomposition}
\label{sec:clifford}

To derive the kinetic equations governing the Clifford components, we expand the Wigner function in the complete basis of the Clifford algebra. The sixteen generators,
\[
\{1,\gamma^5,\gamma^\mu,\gamma^\mu\gamma^5,\sigma^{\mu\nu}\},
\]
form a complete basis for the space of Dirac matrices, where
\(\sigma^{\mu\nu}=\frac{i}{2}[\gamma^\mu,\gamma^\nu]\)
are the generators of the Lorentz group. The Wigner function can therefore be expanded as
\begin{equation}
W = \mathcal{F} + i\gamma^5\mathcal{P} + \gamma^\mu\mathcal{V}_\mu + \gamma^\mu \gamma^5\mathcal{A}_\mu + \frac{1}{2}\sigma^{\mu \nu}\mathcal{S}_{\mu \nu}.
\label{eq:wigner_decomp}
\end{equation}
For convenience, we introduce the function $j_B(z) = \frac{i}{6}j_0 (z) + \frac{1}{2}j_1 (z) - \frac{i}{3}j_2 (z)$, together with the differential operators \(\mathcal Q\) and \(\mathcal D\), defined as $\mathcal{Q} = \left(n\cdot\Pi\right)\left(n\cdot\nabla\right) + \frac{e}{2}j_1(\tfrac{\triangle}{2})\left[\tfrac{n\cdot \partial_x}{2}; n_\mu F^{\mu \nu} \right]\partial^p_\nu$ and $\mathcal{D} = \left(n\cdot\Pi\right)^2 - \left(\frac{n\cdot\nabla}{2}\right)^2 - \left\{\frac{1}{6}j_0(\tfrac{\triangle}{2}) - \frac{1}{3}j_2 (\tfrac{\triangle}{2})\right\}\left[\tfrac{n\cdot \partial_x}{2}; n_\mu F^{\mu \nu} \right]\partial^p_\nu$. Substituting the decomposition into Eq.~\eqref{eq:W_EOM_3} and projecting onto the Clifford basis yields a coupled system of equations for the Clifford components. Since Eq.~\eqref{eq:W_EOM_3} depends on the complex combination \(K^\mu=\Pi^\mu+i\nabla^\mu/2\), the projected equations naturally split into independent real and imaginary parts. The real part is given by
\begin{equation}
\left(\Pi\cdot\mathcal{V} \right) - m\mathcal{F}=\frac{1}{M_{\rm Pl}}\mathcal{D}\,\left\{\eta_1\left(n\cdot\mathcal{V}\right) - \eta_2\left(n\cdot\mathcal{A}\right)\right\},
\label{eq:real:1}
\end{equation}
\begin{equation}
\begin{aligned}
    -\frac{\nabla\cdot\mathcal{A}}{2}+m\mathcal{P}= &-\frac{1}{M_{\rm Pl}}\mathcal{Q}\, \left\{ \eta_1 \left(n\cdot\mathcal{A}\right) - \eta_2 \left(n\cdot\mathcal{V} \right) \right\},
\end{aligned}
\label{eq:real:2}
\end{equation}
\begin{equation}
\begin{aligned}
    \Pi_\mu\mathcal{F} + \frac{\nabla^\nu}{2}\mathcal{S}_{\mu\nu}-m\mathcal{V}_\mu = &\frac{1}{M_{\rm Pl}}\mathcal{D}\,
    \left\{\eta_1n_\mu\mathcal{F} + \eta_2n_\mu\mathcal{P}+2n^\sigma\mathcal{S}_{\sigma\mu}+n^\alpha\tilde{\mathcal{S}}_{\mu\alpha}\right\},
\end{aligned}
\label{eq:real:3}
\end{equation}
\begin{equation}
\begin{aligned}
    -\frac{\nabla_\mu\mathcal{P}}{2}+\Pi^\sigma\tilde{\mathcal{S}}_{\mu\sigma} -m\mathcal{A}_\mu = &-\frac{1}{M_{\rm Pl}}\mathcal{D}\,\left\{\eta_1n_\mu\left(n\cdot\mathcal{V}\right)-\eta_2\left(n_\mu\mathcal{F}+2n^\sigma\mathcal{S}_{\sigma\mu}\right) \right\},
\end{aligned}
\label{eq:real:4}
\end{equation}
\begin{equation}
\begin{aligned}
    &\frac{\nabla_\mu\mathcal{V}_\nu - \nabla_\nu\mathcal{V}_\mu}{2} + \epsilon_{\mu\nu\alpha\beta}\Pi^\alpha\mathcal{V}^\beta - m\mathcal{S}_{\mu\nu} = \\
    &\frac{1}{M_{\rm Pl}}\mathcal{D}\,\left\{ \eta_1\left(g_{\mu\nu}n\cdot\mathcal{V} + n_\nu\mathcal{V}_\mu-n_\mu\mathcal{V}_\nu\right) + n^\alpha\epsilon_{\mu\nu\alpha\sigma}\left(\eta_1\mathcal{A}^\sigma - \eta_2\mathcal{V}^\sigma\right)\right\}.
\end{aligned}
\label{eq:real:5}
\end{equation}
while the imaginary part reads
\begin{equation}
\begin{aligned}
\frac{\nabla\cdot\mathcal{V}}{2} = \frac{1}{M_{\rm Pl}}\mathcal{Q}\, \left\{ \eta_1\left(n\cdot\mathcal{V}\right) - \eta_2\left(n\cdot\mathcal{A}\right) \right\},
\end{aligned}
\label{eq:im:1}
\end{equation}
\begin{equation}
\begin{aligned}
\Pi\cdot\mathcal{A} = \frac{1}{M_{\rm Pl}}\mathcal{D}\,\left\{ \eta_1\left(n\cdot\mathcal{A}\right) - \eta_2\left(n\cdot\mathcal{V}\right) \right\},
\end{aligned}
\label{eq:im:2}
\end{equation}
\begin{equation}
\begin{aligned}
\frac{\nabla_\mu\mathcal{F}}{2} -\Pi^\nu\mathcal{S}_{\mu\nu}= \frac{1}{M_{\rm Pl}}\mathcal{Q}\, \left\{ \eta_1n_\mu\mathcal{F}+\eta_2n_\mu\mathcal{P}+2n^\sigma\mathcal{S}_{\sigma\mu} +n^\alpha\tilde{\mathcal{S}}_{\mu\alpha} \right\},
\end{aligned}
\label{eq:im:3}
\end{equation}
\begin{equation}
\begin{aligned}
\Pi_\mu\mathcal{P} + \frac{1}{2}\nabla^\sigma\tilde{\mathcal{S}}_{\mu\sigma}= -\frac{1}{M_{\rm Pl}}\mathcal{Q}\, \left\{ \eta_1n_\mu\left(n\cdot\mathcal{V}\right)-\eta_2\left(n_\mu\mathcal{F}+2n^\sigma\mathcal{S}_{\sigma\mu}  \right) \right\},
\end{aligned}
\label{eq:im:4}
\end{equation}
\begin{equation}
\begin{aligned}
&\Pi_\nu\mathcal{V}_\mu - \Pi_\mu\mathcal{V}_\nu + \epsilon_{\mu\nu\alpha\beta}\Pi^\alpha\mathcal{V}^\beta = \\
&\frac{1}{M_{\rm Pl}}\mathcal{Q}\, \left\{ \eta_1\left(g_{\mu\nu}n\cdot\mathcal{V} + n_\nu\mathcal{V}_\mu-n_\mu\mathcal{V}_\nu\right) + n^\alpha\epsilon_{\mu\nu\alpha\sigma}\left(\eta_1\mathcal{A}^\sigma - \eta_2\mathcal{V}^\sigma\right) \right\}.
\end{aligned}
\label{eq:im:5}
\end{equation}

The equations above constitute the complete set of coupled quantum transport and constraint equations governing the Clifford components of the Wigner function. Beyond the semiclassical approximation, they provide the starting point for investigating higher-order quantum corrections, spin dynamics, polarization effects, and other quantum phenomena in Lorentz-violating media. In the present work, however, our goal is to establish the relativistic hydrodynamic limit. We therefore restrict the analysis to the semiclassical, spin-unpolarized regime, where the system simplifies considerably. The complete projection of Eq.~\eqref{eq:W_EOM_3} onto the Clifford basis is presented in Appendix~\ref{sec:app1}.

\subsection{Semiclassical limit}
\label{sec:classical}

For the purposes of the present work, we restrict our analysis to the
semiclassical regime, retaining the leading-order contributions in
the expansion in $\hbar$. Restoring the explicit factors of $\hbar$,
the operators introduced in the previous subsection become
\begin{equation}
\nabla^\mu = \hbar\nabla^\mu = \mathcal{O}(\hbar),
\end{equation}
\begin{equation}
\Pi^\mu = p^\mu - \frac{e\hbar}{2}
j_1\!\left(\frac{\hbar\triangle}{2}\right)
F^{\mu\nu}\partial_\nu^p
= p^\mu+\mathcal{O}(\hbar^2).
\end{equation}
The commutator contribution requires particular attention, since
terms of order $\hbar$ may contribute to the kinetic transport
equation. In particular, the Vlasov force arises from the
$\mathcal{O}(\hbar)$ terms in the Wigner equations, despite
corresponding to the leading-order semiclassical dynamics.

The Bessel-function operator entering the commutator contains
both real and imaginary contributions, which enter different
equations in the Clifford decomposition. In the semiclassical
expansion, the contribution to the imaginary equations is
proportional to $j_1(\hbar\triangle/2)$, which vanishes at
leading order. Consequently, the commutator does not contribute
to the leading-order kinetic transport equation. This is demonstrated explicitly in the derivations presented in Appendix~\ref{sec:app:kinectic}.

Furthermore, we restrict our analysis to an unpolarized medium of massive fermions ($m\neq 0$). We assume that no macroscopic spin polarization is present at leading order in the semiclassical expansion, such that $\mathcal{A}_\mu=\mathcal{O}(\hbar)$. The constraint equations of the Wigner formalism for massive fermions relate the pseudoscalar and tensor components to the axial-vector component and its derivatives~\cite{VASAK1987462,Elze:1986qd,Weickgenannt:2019dks}. With this power counting, these relations yield
\begin{equation}
    \mathcal{P},\,\mathcal{A}_\mu,\,\mathcal{S}_{\mu\nu}
    =\mathcal{O}(\hbar).
    \label{eq:spin_suppression}
\end{equation}
Consequently, the semiclassical dynamics can be described solely in terms of the scalar and vector components $\mathcal{F}$ and $\mathcal{V}_\mu$. The resulting cancellation of $\eta_2$ is therefore specific to the unpolarized sector considered here; spin-polarization effects, potentially induced by LIV interactions, lie beyond the scope of the present approximation.

Keeping only the leading-order contributions,
Eq.~\eqref{eq:real:3} reduces to
\begin{equation}
\mathcal{V}_\mu =
\frac{1}{m}
\left[
p_\mu -
\frac{\eta_1}{M_{\rm Pl}}
\left(n\cdot p\right)^2 n_\mu
\right]\mathcal{F}
=
\frac{p_\mu^{\rm LIV}}{m}\mathcal{F},
\label{eq:transport}
\end{equation}
which establishes the algebraic relation between the scalar
and vector components of the Wigner function. Similarly,
substituting Eq.~\eqref{eq:transport} into
Eq.~\eqref{eq:real:1}, we obtain
\begin{equation}
\left[
p_\mu^{\rm LIV}
\left(p^{\rm LIV}\right)^\mu - m^2
\right]\mathcal{F} = 0,
\label{eq:constraint}
\end{equation}
which represents the corresponding mass-shell constraint.

Expanding Eq.~\eqref{eq:constraint} to first order in
$\eta_1/M_{\rm Pl}$ yields
\begin{equation}
p^\mu p_\mu - m^2 -
\frac{2\eta_1}{M_{\rm Pl}}
\left(n\cdot p\right)^3 = 0,
\end{equation}
in agreement with the Lorentz-violating fermionic dispersion
relation obtained directly from the modified Dirac equation.
Thus, the conventional relativistic mass-shell condition
receives Planck-suppressed corrections induced by the
Myers--Pospelov operators.

At leading order, the independent information required to
describe the unpolarized medium is encoded in the scalar
component of the Wigner function. The corresponding
phase-space distribution provides the connection between
the microscopic quantum-field description and the
macroscopic fluid variables.

In Appendix~\ref{sec:app:kinectic}, we further derive the
LIV-modified kinetic transport equation and demonstrate
that its first momentum moment reproduces the macroscopic
energy-momentum balance equation obtained from Noether's
theorem.

\section{Relativistic hydrodynamics of a charged fluid with Lorentz violation}
\label{sec:eqsRMHD}

Having established the semiclassical Wigner relations on the mass shell, we now derive the relativistic hydrodynamic equations for a charged fluid in the presence of Lorentz invariance violation. The macroscopic equations follow from the conservation of the total energy-momentum tensor and of the conserved $U(1)$ current. The fermionic contribution is obtained by taking the ensemble average of the microscopic energy-momentum tensor in terms of the Wigner function, leading to a Lorentz-violating hydrodynamic description of the charged fluid. The electromagnetic contribution is treated classically within the mean-field approximation, so that the total dynamics is described by the coupled evolution of the fluid and electromagnetic field.

For compactness, we formulate the modified field equations and conservation laws in terms of $F^{\mu\nu}$. In the limiting cases discussed in Sec.~\ref{sec:cases}, however, we also express the electromagnetic field tensor and its dual, $\widetilde{F}^{\mu\nu}$, in terms of the electric and magnetic fields, $e^\mu$ and $b^\mu$, respectively.

\subsection{Energy-momentum tensor}

The Myers-Pospelov model introduces a preferred background vector $n^\mu$, whose nature is not fixed a priori. In particular, $n^\mu$ may be timelike, spacelike, or, more generally, spacetime dependent. Throughout this work, we assume $n^\mu$ to be a constant background vector, thereby preserving spacetime translation invariance. According to Noether's theorem, the conserved current associated with this symmetry is the canonical energy-momentum tensor, which can be decomposed into electromagnetic and fermionic contributions,
\begin{equation}
\partial_\mu T_{\rm can}^{\mu\nu} = \partial_\mu
\left( T_{\rm EM}^{\mu\nu} + T_{\rm f}^{\mu\nu} \right) =0.
\end{equation}
This conservation law constitutes one of the fundamental equations governing the relativistic hydrodynamics of the charged fluid. By evaluating the electromagnetic and fermionic sectors separately, we derive the corresponding Lorentz-violating modifications to the macroscopic conservation equations.

\subsubsection{Electromagnetic contribution}

Since the Myers--Pospelov photon sector contains higher-order derivatives of the gauge field, the canonical energy-momentum tensor must be constructed using the generalized Noether prescription for higher-derivative theories. For a Lagrangian depending on the fields $\phi_i$, their first derivatives, and second derivatives, the canonical tensor is given by
\begin{equation}
    T^{\mu\nu}_{\rm can} =
    \frac{\partial\mathcal{L}}{\partial(\partial_\mu\phi_i)}
    \partial^\nu\phi_i
    +
    \frac{\partial\mathcal{L}}{\partial(\partial_\mu\partial_\alpha\phi_i)}
    \partial_\alpha\partial^\nu\phi_i
    -
    \partial_\alpha
    \left(
    \frac{\partial\mathcal{L}}
    {\partial(\partial_\mu\partial_\alpha\phi_i)}
    \right)
    \partial^\nu\phi_i
    -
    g^{\mu\nu}\mathcal L,
    \label{eq:tensorcanonico}
\end{equation}
where $\phi_i$ denotes the QED fields. For the electromagnetic sector, the only relevant dynamical field is the gauge potential, $\phi_i=A^\mu$.

Applying this prescription to the Lorentz-violating photon Lagrangian yields the canonical electromagnetic energy-momentum tensor
\begin{equation}
    \begin{aligned}
        T^{\mu\nu}_{EM} = -F^{\mu\rho}&\partial^\nu A_\rho + \\
        &+\frac{\xi}{M_{\rm Pl}}\left[\frac{3}{2}n^\mu\left(n\cdot\partial\right) \left(n_\theta \tilde{F}^{\theta\rho}\right) - n^\rho\left(n\cdot\partial\right)\left(n_\theta \tilde{F}^{\theta\mu}\right)\right]\partial^\nu A_\rho + \\
        &+\frac{\xi}{M_{\rm Pl}}\left[ n^\sigma F_{\sigma\beta}\frac{n_\theta \epsilon^{\theta\beta\mu\rho}}{2}\left(n\cdot\partial\right) -  \left(n\cdot\partial\right)\left(n^\sigma F_{\sigma\beta}\right) \frac{n_\theta\epsilon^{\theta\beta\mu\rho}}{2} \right]\partial^\nu A_\rho + \\ 
        &+ \frac{\xi}{M_{\rm Pl}}\frac{n^\mu}{2}F_{\sigma\beta}\ n_\theta\ \partial^\nu\tilde{F}^{\theta\beta} - g^{\mu\nu}\mathcal{L}_\gamma.
    \end{aligned}
    \label{eq:tensorEMcan}
\end{equation}
which is neither symmetric nor manifestly gauge invariant. In Lorentz-invariant theories, the first property can be remedied through the Belinfante--Rosenfeld improvement procedure~\cite{BELINFANTE1940449,Rosenfeld1940}. In the present Lorentz-violating framework, although improvement terms remain available, the procedure does not generally yield a fully symmetric energy-momentum tensor due to the explicit breaking of Lorentz symmetry~\cite{Kostelecky:2003fs}.

Since our objective is to derive the conservation equations rather than a symmetric energy-momentum tensor, it is sufficient to evaluate its divergence. Using the generalized Maxwell equations together with the identity
\[
\partial^\nu A_\rho
=
F^\nu{}_\rho+\partial_\rho A^\nu,
\]
we obtain
\begin{equation}
   \begin{aligned}
       \partial_\mu T^{\mu\nu}_{\rm EM} = &-j^\rho_{\rm eff}F^{\nu}{}_\rho - j^\rho_{\rm eff}\partial_\rho A^\nu +\\
       &+\frac{\xi}{M_{\rm Pl}}\left[ \left(n\cdot\partial\right)\left(n_\theta\tilde{F}^{\theta\rho}\right);\partial^\nu\left(n^\sigma F_{\sigma\rho}\right)\right].
   \end{aligned}
   \label{eq:tensorEM}
\end{equation}
The first term corresponds to the Lorentz force density written in terms of the effective current $j_{\rm eff}^\mu$. The commutator arises from the higher-derivative Lorentz-violating operator and is retained at this stage of the derivation. It vanishes upon adopting the Hartree approximation, in which the electromagnetic fields are treated as commuting quantities. The remaining gauge-dependent term, proportional to $j_{\rm eff}^\rho\partial_\rho A^\nu$, is exactly canceled by the corresponding contribution arising from the fermionic sector once the latter is expressed in a gauge-invariant form. Consequently, the total energy-momentum conservation law remains gauge invariant within the adopted approximation.

The algebra leading from Eq.~\eqref{eq:tensorEMcan} to Eq.~\eqref{eq:tensorEM}, as well as the explicit cancellation of the gauge-dependent contribution, is presented in Appendix~\ref{sec:app2}.

To facilitate the hydrodynamic interpretation of the Lorentz-violating force, we decompose the electromagnetic field tensor and its dual in terms of a timelike four-velocity \(u^\mu\), normalized as \(u^\mu u_\mu=1\), and the corresponding comoving electric and magnetic four-vectors, defined by \(e^\mu \equiv F^{\mu\nu}u_\nu\) and \(b^\mu \equiv \tilde{F}^{\mu\nu}u_\nu\), respectively. The four-velocity will be identified with the fluid four-velocity in the following subsection. The resulting decomposition is
\begin{align}
F^{\mu\nu} &= e^\mu u^\nu - e^\nu u^\mu + \epsilon^{\mu\nu\alpha\beta} u_\alpha b_\beta, \label{eq:F_decomp} \\
\tilde{F}^{\mu\nu} &= b^\mu u^\nu - b^\nu u^\mu - \epsilon^{\mu\nu\alpha\beta} u_\alpha e_\beta, \label{eq:Ftilde_decomp}
\end{align}
where $e^\mu u_\mu = 0$ and $b^\mu u_\mu = 0$.

Using the modified inhomogeneous Maxwell equation, Eq.~\eqref{eq:maxwell}, we express the effective current in terms of the electromagnetic field and the LIV current. Substituting this relation into Eq.~\eqref{eq:tensorEM} and employing the comoving-field decomposition, we obtain the divergence of the electromagnetic energy-momentum tensor in terms of \(u^\mu\), \(e^\mu\), and \(b^\mu\):
\begin{equation}
\begin{aligned}
&\partial_\mu T^{\mu\nu}_{\rm EM} = -\left(e^\nu u_\rho - e_\rho u^\nu + \epsilon^\nu{}_\rho{}^{\alpha\beta}u_\alpha b_\beta\right)\partial_\mu \left(e^\mu u^\rho - e^\rho u^\mu + \epsilon^{\mu\rho\alpha\beta}u_\alpha b_\beta\right)\\
&+\frac{2\xi}{M_{\rm Pl}}\left(e^\nu u_\rho - e_\rho u^\nu + \epsilon^\nu{}_\rho{}^{\alpha\beta}u_\alpha b_\beta \right)\left(n\cdot\partial\right)^2\left((n\cdot b)u^\rho - b^ \rho(n\cdot u)-n_\theta\epsilon^{\theta\rho\alpha\beta}u_\alpha e_\beta\right)\\
&+\frac{\xi}{M_{\rm Pl}}\left[\left(n\cdot\partial\right)\left((n\cdot b)u^\rho - b^ \rho(n\cdot u)-n_\theta\epsilon^{\theta\rho\alpha\beta}u_\alpha e_\beta\right);\partial^\nu \left((n\cdot e)u_\rho -e_\rho (n\cdot u)+n^\sigma\epsilon_{\sigma\rho\alpha\beta}u^\alpha b^\beta\right)\right],
\end{aligned}
\label{eq:tensorEM:campos}
\end{equation}
where the gauge-dependent term was dropped as it will be cancelled by the fermionic counterpart.
Equation~\eqref{eq:tensorEM:campos} separates the electromagnetic contribution to the momentum balance into the conventional Lorentz force density and a correction arising from the Lorentz-violating photon sector. The latter emerges when the modified inhomogeneous Maxwell equation is used to express the effective current in terms of the electromagnetic field and the LIV current. The additional commutator term originates from the operator-valued electromagnetic fields in the higher-derivative Lorentz-violating contribution. It vanishes in the adopted Hartree approximation, in which the relevant field products are treated as commuting quantities.

\subsubsection{Fermionic contribution}

Applying the generalized Noether procedure to the fermionic sector, and using Eq.~\eqref{eq:tensorcanonico} with $\phi_i=\psi,\bar{\psi}$, we obtain
\begin{equation}
    \begin{aligned}
        T^{\mu\nu}_{\rm f} =
        \frac{i}{2}\bar{\psi}\gamma^\mu\overleftrightarrow{\partial^\nu}\psi
        +
        \frac{n^\mu}{2M_{\rm Pl}}
        \bar{\psi}\mathcal{H}
        \left[
        \left(n\cdot \overleftrightarrow{D}\right)
        \overleftrightarrow{\partial^\nu}
        \right]
        \psi,
    \end{aligned}
\end{equation}
where $\overleftrightarrow{\partial^\mu}=\partial^\mu-\overleftarrow{\partial^\mu}$.

As for the electromagnetic sector, the canonical energy-momentum tensor is not manifestly gauge invariant. It can therefore be rewritten in terms of gauge-invariant derivatives as
\begin{equation}
    \begin{aligned}
        T^{\mu\nu}_{\rm f} &=
        \frac{i}{2}\bar{\psi}\gamma^\mu\overleftrightarrow{D^\nu}\psi
        +
        \frac{n^\mu}{2M_{\rm Pl}}
        \bar{\psi}\mathcal{H}
        \left[
        \left(n\cdot \overleftrightarrow{D}\right)
        \overleftrightarrow{D^\nu}
        \right]
        \psi
        \\
        &\qquad
        +j_{\rm eff}^\mu A^\nu,
    \end{aligned}
    \label{eq:fermionico}
\end{equation}
where the gauge-dependent term exactly cancels the corresponding contribution appearing in Eq.~\eqref{eq:tensorEM}. Consequently, the total energy-momentum conservation law is gauge invariant.

In order to connect the microscopic field-theoretical description with relativistic hydrodynamics, we express the ensemble average of the fermionic energy-momentum tensor in terms of the covariant Wigner function introduced in Sec.~\ref{sec:wigner}. Using the inverse Fourier transform of the Wigner function,
\begin{equation}
    \left<
    :
    \bar{\psi}(x_+)
    U(A,x_+,x_-)
    \psi(x_-)
    :
    \right>
    =
    \int d^4p\,
    e^{ip\cdot y}
    W(x,p),
\end{equation}
the macroscopic fermionic energy-momentum tensor becomes
\begin{equation}
\begin{aligned}
    T^{\mu\nu}_{\rm fluid}
    =
    \left<
    T^{\mu\nu}_{\rm f}
    \right>
    =
    {\rm tr}
    \int d^4p\,
    e^{ip\cdot y}
    \left[
    p^\nu\gamma^\mu W
    -
    \frac{2}{M_{\rm Pl}}
    (n\cdot p)
    n^\mu
    p^\nu
    \mathcal{H}W
    \right]
    \Bigg|_{y=0}.
\end{aligned}
\end{equation}
Performing the Dirac trace and using Eq.~\eqref{eq:transport}, the tensor can be expressed entirely in terms of the scalar component of the Wigner function,
\begin{equation}
\begin{aligned}
    T^{\mu\nu}_{\rm fluid}
    =
    4
    \int d^4p
    \left[
    p^\mu p^\nu
    -
    \frac{3\eta_1}{M_{\rm Pl}}
    (n\cdot p)^2
    n^\mu
    p^\nu
    \right]
    \frac{\mathcal{F}}{m}.
\end{aligned}
\label{eq:integral:tensor}
\end{equation}
In the semiclassical limit, the scalar component of the Wigner function is given by
\begin{equation}
 \mathcal{F}
 =\frac{1}{(2\pi)^3}
 m\,
 \theta(p^0)\,
 \delta\!\left(
 p_\mu^{\rm LIV}
 (p^{\rm LIV})^\mu
 -
 m^2
 \right)
 f(x,p),
 \label{eq:Fmathcal}
\end{equation}
where $\theta(x)$ denotes the Heaviside function and $f(x,p)$ is the classical phase-space distribution function \cite{VASAK1987462,Elze:1987yb}. The Dirac delta is modified according to the Lorentz-violating mass-shell constraint given by Eq.~\eqref{eq:constraint}.

As shown in Appendix~\ref{sec:app3}, the modified mass-shell delta function can be expanded perturbatively to first order in the Lorentz-violating parameter. After performing the integration over $p^0$, we adopt a local Fermi--Dirac distribution,
\begin{equation}
f(u\cdot p)=\left[e^{\frac{u\cdot p-\mu}{T}}+1\right]^{-1},
\end{equation}
as a closure assumption for the semiclassical kinetic description. This choice parametrizes the local state in terms of the temperature, chemical potential, and fluid four-velocity, but should not be interpreted as an equilibrium solution derived from the collisionless (Hartree) transport equations. In the presence of Lorentz invariance violation, the preferred background may modify the equilibrium distribution through the modified dispersion relations and, for a spacelike background, introduce directional dependence in momentum space~\cite{Colladay:2004qt,Potting:2022fyw}. The derivation of the most general equilibrium distribution compatible with the Lorentz-violating dynamics is beyond the scope of the present work.

With this assumption, the fermionic energy-momentum tensor becomes
\begin{equation}
    \begin{aligned}
        T^{\mu\nu}_{\rm fluid}
        =
        2
        \int
        \frac{d^3p}{(2\pi)^3(u\cdot p)}
        f(u\cdot p)
        \Bigg\{
        p^\mu p^\nu
        +
        \frac{\eta_1}{M_{\rm Pl}}
        (n\cdot p)^2
        \Bigg[
        \frac{3(u\cdot n)p^\mu p^\nu}{u\cdot p}
        -
        \frac{(n\cdot p)p^\mu p^\nu}{(u\cdot p)^2}
        \\
        +
        \frac{1}{f(u\cdot p)}
        \frac{\partial f}{\partial(u\cdot p)}
        \frac{(n\cdot p)p^\mu p^\nu}{u\cdot p}
        -
        3n^\mu p^\nu
        \Bigg]
        \Bigg\}.
    \end{aligned}
    \label{eq:int_covariant_fluid}
\end{equation}
To express Eq.~\eqref{eq:int_covariant_fluid} in the familiar perfect-fluid form, we introduce the Israel--Stewart thermodynamic integrals,
\begin{equation}
I_{nq}
=
\frac{4\pi}{(2q+1)!!}
\int \frac{dp}{(2\pi)^3}\,
\frac{p^{2}}{u\cdot p}\,
f(u\cdot p)\,
(u\cdot p)^{\,n-2q}
\left(\bar p^\mu\bar p_\mu\right)^q,
\label{eq:Israel}
\end{equation}
where $\Delta^{\mu\nu}=g^{\mu\nu}-u^\mu u^\nu$, and $\bar p^\mu=\Delta^\mu_{\ \nu}p^\nu$. Since $\bar p^\mu$ is purely spatial in the local rest frame,
$\bar p^\mu\bar p_\mu=-p^2$, so that
$[\bar p^\mu\bar p_\mu]^q= \left( -1\right)^{q}p^{2q}$ reproduces the standard
Israel--Stewart thermodynamic integrals \cite{ISRAEL1979341}.

The most general second-rank tensor compatible with the available four-vectors $u^\mu$ and $n^\mu$ is
\begin{equation}
    T^{\mu\nu}
    =
    Au^\mu u^\nu
    +
    B\Delta^{\mu\nu}
    +
    C\bar n^\mu\bar n^\nu
    +
    D_1u^\mu\bar n^\nu
    +
    D_2u^\nu\bar n^\mu,
\end{equation}
where $\bar n^\mu = n^\mu-(u\cdot n)u^\mu$ \cite{Romatschke_Romatschke_2019}.
Since the canonical energy-momentum tensor is not required to be symmetric, the coefficients $D_1$ and $D_2$ are treated as independent. Projecting Eq.~\eqref{eq:int_covariant_fluid} onto the tensor basis above, as detailed in Appendix~\ref{sec:app3}, determines the coefficients
$A$, $B$, $C$, $D_1$, and $D_2$ exactly.
In order to identify the corresponding thermodynamic variables, we subsequently adopt the Landau frame, in which the fluid four-velocity is defined as the timelike eigenvector of the energy-momentum tensor \cite{landau1987fluid}. Within this frame, the coefficients are naturally interpreted as Lorentz-violating corrections to the equilibrium energy density and pressure, yielding
\begin{equation}
\begin{aligned}
T^{\mu\nu}_{\rm fluid}
=
(\epsilon_{\rm LIV}+P_{\rm LIV})
u^\mu u^\nu
-
P_{\rm LIV}g^{\mu\nu}
+\frac{\eta_1}{M_{\rm Pl}}
\left[
3(u\!\cdot\!n)^2I_{30}
+
3(\bar n\!\cdot\!\bar n)I_{31}
\right]
u^\mu\bar n^\nu,
\end{aligned}
\label{eq:fluidLIV}
\end{equation}
where
\begin{equation}
P_{\rm LIV}
=
P_0
-
\frac{\eta_1}{M_{\rm Pl}}
\left[
(u\!\cdot\!n)^3I_{30}
+
3(u\!\cdot\!n)
(\bar n\!\cdot\!\bar n)
I_{31}
\right],
\end{equation}
and
\begin{equation}
\begin{aligned}
\epsilon_{\rm LIV}
=
&\epsilon_0
+
\frac{\eta_1}{M_{\rm Pl}}
\Big[
(u\!\cdot\!n)^3I_{3,-1}
-
3(u\!\cdot\!n)^3I_{30}
+
3(u\!\cdot\!n)
(\bar n\!\cdot\!\bar n)
I_{30}
+
7(u\!\cdot\!n)
(\bar n\!\cdot\!\bar n)
I_{31}
\Big].
\end{aligned}
\end{equation}

Equation~\eqref{eq:fluidLIV} represents the Lorentz-violating energy-momentum tensor of the charged fluid. Its structure reveals two qualitatively distinct effects induced by Lorentz violation. The first consists of a renormalization of the effective thermodynamic variables, preserving the standard perfect-fluid structure. The second originates from the preferred background vector and gives rise to an anisotropic, nonsymmetric contribution to the energy-momentum tensor. As discussed in Sec.~\ref{sec:cases}, the physical manifestation of these terms depends on whether the Lorentz-violating background is timelike or spacelike.

In the standard formulation of General Relativity, the energy--momentum tensor is symmetric. In the present framework, LIV induces an nonsymmetric contribution to the energy--momentum tensor, preventing its direct incorporation into the conventional GRMHD framework. A consistent extension to curved spacetime would require a suitable treatment of the energy--momentum tensor and the Lorentz-violating background, following the general considerations discussed in Ref.~\cite{Kostelecky:2003fs}. The present formalism is therefore restricted to flat spacetime.

\subsubsection{Conservation equation}

Using the conservation of the total energy-momentum tensor
and combining the contributions from the electromagnetic
and fermionic sectors, we obtain
\begin{equation}
\partial_\mu T^{\mu\nu}_{\rm fluid}
=
j^{\mu}_{\rm eff}F^\nu{}_\mu,
\label{eq:total_conservation}
\end{equation}
where $T^{\mu\nu}_{\rm fluid}$ is given by
Eq.~\eqref{eq:fluidLIV}. The commutator contribution
vanishes within the c-number/Hartree approximation,
while the gauge-dependent terms cancel exactly between
the electromagnetic and fermionic sectors, as explicitly
demonstrated in Appendix~\ref{sec:app2}.

Equation~\eqref{eq:total_conservation} describes the
energy-momentum balance of the relativistic charged
fluid in the presence of Myers--Pospelov Lorentz
invariance violation. As an independent consistency
check, Appendix~\ref{sec:app:kinectic} derives the
corresponding LIV-modified kinetic transport equation
and demonstrates that its first momentum moment
reproduces Eq.~\eqref{eq:total_conservation}, in
agreement with the conservation law obtained from
Noether's theorem.

\subsection{Particle conservation}
\label{sub:sec:u1}

The Myers--Pospelov effective theory preserves local $U(1)$ gauge invariance despite the presence of Lorentz-violating operators. Consequently, the theory still possesses a conserved Noether current associated with the global phase symmetry. Using the generalized Noether formalism for higher-derivative theories, the corresponding current is given by
\begin{equation}
\begin{aligned}
 J^\mu_{\rm Noether} = \frac{\partial\mathcal{L}}{\partial(\partial_\mu\psi)}\psi + \frac{\partial\mathcal{L}}{\partial(\partial_\mu\partial_\nu\psi)}\partial_\nu\psi - \partial_\nu\left(\frac{\partial\mathcal{L}}{\partial(\partial_\mu\partial_\nu\psi)}\right)\psi \\
 -\bar{\psi}\frac{\partial\mathcal{L}}{\partial(\partial_\mu\bar\psi)} -\partial_\nu\bar\psi\frac{\partial\mathcal{L}}{\partial(\partial_\mu\partial_\nu\bar\psi)} +\bar\psi \partial_\nu\left(\frac{\partial\mathcal{L}}{\partial(\partial_\mu\partial_\nu\psi)}\right).
\end{aligned}
\label{eq:u1}
\end{equation}
Applying Eq.~\eqref{eq:u1} to the Myers--Pospelov Lagrangian yields
\begin{equation}
    J^{\mu}_{\rm Noether} = j^\mu - \frac{ie}{M_{\rm Pl}} n^\mu \bar{\psi}\left(\eta_1 \slashed{n}+\eta_2 \slashed{n}\gamma^5\right)\left(n \cdot \overleftrightarrow{D}\right)\psi = j^\mu_{\rm eff}.
    \label{eq:current}
\end{equation}
which coincides with the effective current appearing in the modified Maxwell equations. Therefore, charge conservation is preserved, 
\begin{equation}
    \partial_\mu j_{\rm eff}^\mu = 0,
    \label{eq:conservacorrente}
\end{equation}
although the conserved current acquires Lorentz-violating corrections.

In order to obtain the corresponding hydrodynamic current, we again express the ensemble average in terms of the Clifford components of the Wigner function. 
\begin{equation}
    \langle j^\mu_{\rm eff} \rangle
    =
    4e
    \int d^4p
    \left[
        p^\mu
        -
        3\frac{\eta_1}{M_{\rm Pl}}
        (n\cdot p)^2 n^\mu
    \right]
    \frac{\mathcal{F}}{m}.
    \label{eq:integral:corrente}
\end{equation}
Following the same procedure employed for the energy-momentum tensor, we obtain
\begin{equation}
    \begin{aligned}
        \left< j^\mu_{\rm eff}\right> = 4e
        \int
        \frac{d^3p}{2(u\cdot p)}
        f(u\cdot p)
        \Bigg\{
        p^\mu
        +
        \frac{\eta_1}{M_{\rm Pl}}
        (n\cdot p)^2
        \Bigg[
        \frac{3(u\cdot n)p^\mu}{u\cdot p}
        -
        \frac{(n\cdot p)p^\mu}{(u\cdot p)^2}
        \\
        +
        \frac{1}{f(u\cdot p)}\frac{\partial f}{\partial(u\cdot p)}
        \frac{(n\cdot p)p^\mu}{u\cdot p}
        -
        3n^\mu
        \Bigg]
        \Bigg\}.  
    \end{aligned}
\end{equation}
The most general decomposition of the conserved current compatible with the available four-vectors is
\begin{equation}
    J^\mu = Au^\mu + B\bar{n}^\mu.
\end{equation}
The coefficients are obtained by projecting the covariant integral onto this basis. Their explicit derivation is presented in Appendix~\ref{sec:app3}. As in the previous subsection, we finally adopt the Landau frame, allowing the coefficients to be interpreted as corrections to the equilibrium particle density and to an anisotropic particle flux. The hydrodynamic conserved current then takes the form
\begin{equation}
    \partial_\mu\left(\rho_{\rm LIV}u^\mu + \mathcal{C}\bar{n}^\mu \right) = 0,
\end{equation}
where
\begin{equation}
    \begin{aligned}
        \rho_{\rm LIV}
        =
        \rho_0
        +
        \frac{\eta_1}{M_{\rm Pl}}
        \Big[
        &
        (u\cdot n)^3I_{2,-1}
        -
        4(u\cdot n)^3I_{20}
        \\
        &
        -
        6(u\cdot n)
        (\bar n\cdot\bar n)I_{21}
        +
        3(u\cdot n)
        (\bar n\cdot\bar n)I_{20}
        \Big],
    \end{aligned}
\end{equation}
and
\begin{equation}
    \begin{aligned}
        \mathcal{C}
        =
        \frac{\eta_1}{M_{\rm Pl}}
        \left(\frac{\rho_0}{P_0+\epsilon_0}\right)
        \left[
            3(\bar n\cdot\bar n)I_{32}
            +
            9(u\cdot n)^2I_{31}
        \right].
    \end{aligned}
\end{equation}

The Lorentz-violating corrections exhibit the same qualitative structure found for the energy-momentum tensor. The equilibrium particle density is renormalized by Planck-suppressed contributions, while an additional particle flux is induced along the preferred background direction. As in the previous case, the relative importance of these corrections depends on the nature of the Lorentz-violating background.

\subsection{Equation of state}

In practical applications, the hydrodynamic equations must be supplemented by an equation of state. Since the Lorentz-violating corrections derived above are expressed in terms of the thermodynamic integrals $I_{nq}$, it is convenient to rewrite them in terms of the corresponding Lorentz-invariant thermodynamic variables. In this way, any existing equation of state can be straightforwardly generalized to include first-order Lorentz-violating effects.

To relate the physical thermodynamic variables in the Lorentz-violating theory to those of the corresponding Lorentz-invariant system, we introduce the auxiliary quantities $T_0$ and $\mu_0$, defined through
\begin{equation}
    T=T_0+\delta T,
\end{equation}
\begin{equation}
    \mu=\mu_0+\delta\mu,
\end{equation}
where the corrections $\delta T$ and $\delta\mu$ are of order
$\mathcal{O}(M_{\rm Pl}^{-1})$.

We further adopt the Landau matching conditions,
\begin{equation}
    \epsilon_{\rm LIV}(T,\mu)
    =
    \epsilon_0(T_0,\mu_0),
    \label{eq:mathclandau1}
\end{equation}
and
\begin{equation}
    \rho_{\rm LIV}(T,\mu)
    =
    \rho_0(T_0,\mu_0),
    \label{eq:mathclandau2}
\end{equation}
which define the auxiliary equilibrium variables.

The Lorentz-violating thermodynamic quantities can be written as
\begin{equation}
    \epsilon_{\rm LIV}(T,\mu)
    =
    \epsilon_0(T,\mu)
    +
    \Delta\epsilon(T,\mu),
\end{equation}
\begin{equation}
    \rho_{\rm LIV}(T,\mu)
    =
    \rho_0(T,\mu)
    +
    \Delta\rho(T,\mu),
\end{equation}
where $\Delta\epsilon$ and $\Delta\rho$ denote the explicit first-order Lorentz-violating corrections.

Expanding $\epsilon_0(T,\mu)$ and $\rho_0(T,\mu)$ around $(T_0,\mu_0)$, imposing the matching conditions
Eqs.~\eqref{eq:mathclandau1} and \eqref{eq:mathclandau2}, and retaining terms up to first order in $M_{\rm Pl}^{-1}$, yields
\begin{equation}
\begin{cases}
\Delta\epsilon
+
\displaystyle
\frac{\partial\epsilon_0}{\partial T}\delta T
+
\frac{\partial\epsilon_0}{\partial\mu}\delta\mu
=0,
\\[0.3cm]
\Delta\rho
+
\displaystyle
\frac{\partial\rho_0}{\partial T}\delta T
+
\frac{\partial\rho_0}{\partial\mu}\delta\mu
=0.
\end{cases}
\end{equation}
Solving this system gives
\begin{equation}
    \delta T
    =
    \frac{1}{\mathcal J}
    \left(
    \frac{\partial\epsilon_0}{\partial\mu}\Delta\rho
    -
    \frac{\partial\rho_0}{\partial\mu}\Delta\epsilon
    \right),
\end{equation}
\begin{equation}
    \delta\mu
    =
    \frac{1}{\mathcal J}
    \left(
    \frac{\partial\rho_0}{\partial T}\Delta\epsilon
    -
    \frac{\partial\epsilon_0}{\partial T}\Delta\rho
    \right),
\end{equation}
where
\begin{equation}
\mathcal J
=
\frac{\partial\rho_0}{\partial T}
\frac{\partial\epsilon_0}{\partial\mu}
-
\frac{\partial\rho_0}{\partial\mu}
\frac{\partial\epsilon_0}{\partial T}.
\end{equation}

This construction allows any thermodynamic quantity in the Lorentz-violating theory to be expressed directly in terms of the corresponding Lorentz-invariant equation of state. In particular, expanding the pressure around $(T_0,\mu_0)$ gives
\begin{equation}
\begin{aligned}
P_{\rm LIV}(T,\mu)
=
P_0(T_0,\mu_0)
&+
\frac{\partial P_0}{\partial T}\,\delta T
+
\frac{\partial P_0}{\partial\mu}\,\delta\mu
+
\Delta P,
\end{aligned}
\end{equation}
where $\delta T$ and $\delta\mu$ are given by the previous equations.

Finally, since every Lorentz-violating correction is explicitly suppressed by the Planck scale, evaluating the thermodynamic integrals at $(T,\mu)$ or $(T_0,\mu_0)$ only differs by higher-order terms. Therefore, to first order,
\begin{equation}
    \frac{\eta_1}{M_{\rm Pl}}
    I_{ab}(T,\mu)
    \simeq
    \frac{\eta_1}{M_{\rm Pl}}
    I_{ab}(T_0,\mu_0).
\end{equation}

\section{Limiting cases of the background vector}
\label{sec:cases}

The equations derived in the previous section provide a covariant relativistic fluid framework incorporating Myers--Pospelov Lorentz invariance violation for general fluid four-velocity and background vectors. To isolate the physical effects associated with the temporal and spatial components of the background, we consider two highly idealized limiting configurations.

In the first case, the background is purely timelike and locally aligned with the fluid four-velocity, $n^\mu=u^\mu$. We further restrict our analysis to special flow configurations for which the fluid four-velocity is constant, $\partial_\mu u^\nu=0$, throughout the region considered. In the second case, the background is purely spacelike in the local fluid rest frame, satisfying $u\cdot n=0$. Here, we restrict our analysis to flows that preserve this orthogonality condition throughout the region considered.

These limiting configurations are not intended to describe generic flows, but rather to isolate the distinct contributions of the temporal and spatial components of the LIV background. The following subsections examine their respective implications for the electromagnetic and hydrodynamic sectors.

\subsection{Temporal background vector}

For a purely timelike background, the preferred four-vector is aligned with the fluid four-velocity, $n^\mu=u^\mu$. In this limit, the modified hydrodynamic equations for the charged fluid reduce to
\begin{equation}
    \partial_\mu\left(b^\mu u^\nu - b^\nu u^\mu + \epsilon^{\mu\nu\alpha\beta}u_\alpha e_\beta\right) = 0,
\end{equation}
\begin{equation}
    \partial_\mu F^{\mu\nu}
    =
    j_{\rm eff}^\nu
    -
    \frac{2\xi}{M_{\rm Pl}}
    \left(u \cdot \partial\right)^2b^\nu,
    \label{eq:max:temp}
\end{equation}
\begin{equation}
    \partial_\mu T^{\mu\nu}_{\rm fluid}
    =
    j^\rho_{\rm eff}F^\nu{}_\rho,
\end{equation}
\begin{equation}
    T^{\mu\nu}_{\rm fluid}
    =
    \left[
    \epsilon_0
    +
    P_0
    +
    \frac{\eta_1}{M_{\rm Pl}}
    \left(
    I_{3,-1}
    -
    4I_{30}
    \right)
    \right]
    u^\mu u^\nu
    -
    \left[
    P_0
    -
    \frac{\eta_1}{M_{\rm Pl}}I_{30}
    \right]
    g^{\mu\nu},
\end{equation}
\begin{equation}
    \partial_\mu
    \left\{
    \left[
    \rho_0
    +
    \frac{\eta_1}{M_{\rm Pl}}
    \left(
    I_{2,-1}
    -
    4I_{20}
    \right)
    \right]
    u^\mu
    \right\}
    =
    0.
\end{equation}
The electromagnetic field tensor \(F^{\mu\nu}\) is expressed in terms of the comoving variables \(u^\mu\), \(b^\mu\), and \(e^\mu\) through the decomposition given in Eq.~\eqref{eq:F_decomp}.

The hydrodynamic structure of the charged-fluid equations remains unchanged. Lorentz-violating effects appear exclusively through Planck-suppressed corrections to the thermodynamic variables, giving rise to effective energy density, pressure, and particle density. The electromagnetic sector is modified only through the additional higher-time-derivative Myers--Pospelov term in Maxwell's equations. Therefore, a purely timelike background preserves the perfect-fluid form of the fluid energy-momentum tensor, with Lorentz violation acting primarily through corrections to the effective thermodynamic variables and the equation of state.

\subsection{Spatial background vector}

For a purely spacelike background, the preferred vector takes the form
$n^\mu=(0,\hat n)$ in the fluid rest frame. In this limit, the modified hydrodynamic equations for the charged fluid reduce to
\begin{equation}
    \partial_\mu\left(b^\mu u^\nu - b^\nu u^\mu + \epsilon^{\mu\nu\alpha\beta}u_\alpha e_\beta\right) = 0,
\end{equation}
\begin{equation}
\partial_\mu F^{\mu\nu}
=
j_{\rm eff}^\nu
+
\frac{2\xi}{M_{\rm Pl}}
(\hat n\!\cdot\!\nabla)^2
\left(
n_i\tilde{F}^{i\nu}
\right),
\label{eq:max:spatial}
\end{equation}
\begin{equation}
\begin{aligned}
\partial_\mu T^{\mu\nu}_{\rm fluid}
=
j^\rho_{\rm eff}F^\nu{}_\rho,
\end{aligned}
\end{equation}
\begin{equation}
\begin{aligned}
T^{\mu\nu}_{\rm fluid}
=
(\epsilon_0+P_0)
u^\mu u^\nu
-
P_0g^{\mu\nu}
-
\frac{3\eta_1}{M_{\rm Pl}}
I_{31}
u^\mu\bar n^\nu,
\end{aligned}
\end{equation}
\begin{equation}
\partial_\mu
\left\{
\rho_0u^\mu
-
\frac{3\eta_1}{M_{\rm Pl}}
\left( \frac{\rho_0}{P_0+\epsilon_0}\right)I_{32}
\, \bar n^\mu
\right\}
=
0.
\end{equation}

Unlike the purely timelike case, the effective thermodynamic variables
remain unchanged in this limit. Instead, Lorentz violation manifests
itself in both the electromagnetic and fluid sectors through the
preferred spatial direction. In Maxwell's equations, the Myers--Pospelov
correction becomes proportional to $(\hat n\!\cdot\!\nabla)^2$,
introducing anisotropic modifications to the electromagnetic dynamics along the background
direction. In the hydrodynamic sector, the preferred vector generates
new anisotropic contributions to both the fluid energy-momentum tensor
and the particle current. In particular, an additional particle flux appears along $\bar n^\mu$, while the nonsymmetric contribution to the energy-momentum tensor reflects the explicit breaking of rotational invariance by the spacelike background. This implies that the fluid angular momentum is not independently conserved, resulting in an effective torque density exerted by the preferred direction on the charged fluid.

These two limiting cases illustrate how the same covariant formulation leads to qualitatively different phenomenology depending on the nature of the Lorentz-violating background. Generic configurations with both temporal and spatial components interpolate continuously between these two regimes.

\section{Order-of-magnitude estimates and phenomenological outlook}
\label{sec:estimates}

The Lorentz-violating corrections derived in this work are suppressed by the Planck scale. In this section, we establish their characteristic scales using natural units and adopting $\xi, \eta_1 = \mathcal{O}(1)$ as benchmark theoretical parameters. These estimates are perturbative and hold within the effective field theory regime of validity, where $|\xi| k / M_{\rm Pl} \ll 1$ and $|\eta_1| \mathcal{E} / M_{\rm Pl} \ll 1$, with $k$ denoting the characteristic wave number and $\mathcal{E} \sim \max(T,\mu)$ the fluid rest-frame particle energy.

As a first step toward a full magnetohydrodynamic description, we estimate the electromagnetic corrections for magnetically dominated configurations, using $B$ as the characteristic magnetic-field strength. This estimate does not impose the ideal-conductivity condition, whose compatibility with the Lorentz-violating dynamics remains to be investigated \cite{Koide:2009yx}.

\subsection{Electromagnetic sector}

In the electromagnetic sector, the Lorentz-violating force density $f_\xi^\nu$ can be extracted from Eq.~\eqref{eq:tensorEM} in the Hartree limit, yielding a force magnitude of
\begin{equation}
    |f_\xi| \sim \frac{|\xi|}{M_{\rm Pl}}\frac{B^2}{L_n^2},
\end{equation}
where $L_n$ denotes the gradient scale along the preferred background direction. Comparing this correction with the conventional Lorentz force density $|f_{\rm Lorentz}| \sim B^2 / L_J$ over a characteristic current scale $L_J \sim L_n \sim L$ gives
\begin{equation}
    \frac{|f_\xi|}{|f_{\rm Lorentz}|}
    \sim \frac{|\xi|}{M_{\rm Pl} L}
    \sim \frac{|\xi| k}{M_{\rm Pl}}
    \simeq 1.6 \times 10^{-43} |\xi|
    \left(\frac{10^{10}\,\mathrm{cm}}{L}\right).
    \label{eq:ratio_xi_normalized}
\end{equation}
For macroscopic astrophysical scales such as $L = 10^{10}\,\mathrm{cm}$ or $10^{15}\,\mathrm{cm}$ \cite{Meszaros:2006rc,Piran:2004ba}, the ratio is suppressed to approximately $1.6 \times 10^{-43}|\xi|$ and $1.6 \times 10^{-48}|\xi|$, respectively. The correction becomes comparable to the conventional force only for $L \sim |\xi|/M_{\rm Pl}$, which, for $\xi \sim \mathcal{O}(1)$, corresponds to the Planck length. Such scales lie outside the regime of validity of the effective theory, as they violate the perturbative condition $|\xi|k/M_{\rm Pl}\ll1$.

Observational constraints on the photon sector, derived primarily from gamma-ray burst (GRB) polarimetry and vacuum birefringence, bound the coupling to $|\xi| \lesssim 10^{-14}\text{--}10^{-16}$~\cite{Laurent:2011he,Gotz:2014vza,Toma:2012xa}. Combining these empirical bounds with Eq.~\eqref{eq:ratio_xi_normalized}, the resulting force ratio is bounded by approximately $10^{-57}\text{--}10^{-59}$ for $L = 10^{10}\,\mathrm{cm}$, and $10^{-62}\text{--}10^{-64}$ for $L = 10^{15}\,\mathrm{cm}$. Thus, the photon-sector correction is negligible compared with the conventional Lorentz force on the macroscopic scales considered here.

\subsection{Fermionic sector}

In the fermionic sector, the relative Lorentz-violating corrections to the background energy density $\epsilon_0$, pressure $P_0$, and particle density $\rho_0$ follow a unified scaling in the relativistic regime ($\mathcal{E} \gg m$):
\begin{equation}
    \frac{\delta X}{X_0} \sim \eta_1 \frac{\mathcal{E}}{M_{\rm Pl}},
    \qquad X \in \{\epsilon,P,\rho\},
    \label{eq:ratio_eta_generic}
\end{equation}
while the anisotropic flux follows the same suppression pattern and can be estimated separately as
\begin{equation}
    \frac{|C\bar n|}{\rho_0}
    \sim \eta_1\frac{\mathcal{E}}{M_{\rm Pl}}.
\end{equation}

For hot astrophysical environments, the dimensionless thermal suppression evaluates to
\begin{equation}
    \frac{\mathcal{E}}{M_{\rm Pl}}
    \simeq 8.2 \times 10^{-20}
    \left(\frac{\mathcal{E}}{1\,\mathrm{GeV}}\right),
    \label{eq:eta_normalized}
\end{equation}
yielding $\mathcal{E}/M_{\rm Pl} \sim 8.2 \times 10^{-23}$ at $\mathcal{E} = 1\,\mathrm{MeV}$ and $8.2 \times 10^{-21}$ at $\mathcal{E} = 100\,\mathrm{MeV}$, characteristic of hot astrophysical environments such as gamma-ray burst engines and neutron-star merger remnants \cite{Vurm:2011fq,Peer:2015eek,Faber:2012rw}.

Empirical bounds on the dimension-five fermionic operators come from
Crab Nebula observations. The synchrotron emission constrains
$|\eta_\pm|\lesssim 10^{-5}$ at 95\% C.L.~\cite{Maccione:2007}, where
$\eta_{R,L}=\eta_1\pm\eta_2$, while for the sign that lowers the maximum
electron speed the limit is $\eta\gtrsim -7\times10^{-8}$
\cite{Jacobson:2003}. Adopting $|\eta_1|\lesssim 10^{-5}$ and
$\mathcal{E}$ between $1$ and $100\,\mathrm{MeV}$, the relative
corrections are bounded by
\begin{equation}
    \frac{\delta X}{X_0}\lesssim 8\times10^{-28}\text{--}8\times10^{-26}.
\end{equation}
The benchmark value $\eta_1 = \mathcal{O}(1)$ used initially serves purely to illustrate the natural scale of the operator; realistic evaluations must incorporate these empirical bounds and the naturalness considerations discussed in Sec.~\ref{sec:liv}.

\subsection{Dynamical amplification and outlook}

Although local background corrections to smooth equilibria are negligible, the derived relativistic hydrodynamic equations remain nonlinear. In particular, a spacelike background introduces qualitatively new anisotropic effects, including a preferred-direction particle flux and a torque density associated with the nonsymmetric energy-momentum tensor. Together with vacuum birefringence in the photon sector, these contributions could potentially act as symmetry-breaking seeds in the dynamics of astrophysical plasmas. In flows subject to instabilities, such as Kelvin--Helmholtz~\cite{Bucciantini:2006jg,Mizuno:2006ma} or Rayleigh--Taylor modes~\cite{Meliani:2007su,Meliani:2009vq}, or magnetic reconnection~\cite{Sironi:2014jfa}, these effects could in principle be amplified. Similarly, small modifications to the dynamical coefficients or propagation speeds could become relevant for flows sufficiently close to a stability threshold~\cite{Meliani:2008cz}. These possibilities remain speculative and require a dedicated stability analysis.

Determining whether such mechanisms can produce observable signatures requires a complete $3+1$ decomposition and nonlinear numerical simulations, which are left for future work. In particular, establishing which Lorentz-violating contributions survive in an ideal-RMHD description, and under what conditions, is an important next step. The equations derived here provide a starting point for investigating these questions.

\section{Conclusions}
\label{sec:conc}

In this work, we derived the relativistic hydrodynamic equations for a charged fluid incorporating Lorentz invariance violation within the Myers--Pospelov effective field theory. Starting from the modified QED Lagrangian, we derived the energy-momentum tensor and the conserved $U(1)$ current associated with the fermionic sector. By combining these results with the modified Maxwell equations and employing the covariant Wigner formalism, we established the corresponding macroscopic fluid description.

The resulting hydrodynamic equations incorporate Planck-suppressed corrections in both the electromagnetic and matter sectors. In particular, the fluid energy-momentum tensor acquires corrections to the thermodynamic quantities, together with additional anisotropic contributions associated with the preferred background vector. Likewise, the particle current receives Lorentz-violating corrections that can give rise to anisotropic particle transport. These modifications reveal how the microscopic Lorentz-violating dynamics affect the macroscopic behavior of a relativistic charged fluid.

The analysis of the limiting cases reveals two qualitatively distinct regimes. For a purely timelike background, the Lorentz-violating corrections can be absorbed into effective thermodynamic quantities, preserving the isotropic structure of the fluid equations. In contrast, a purely spacelike background introduces explicitly direction-dependent contributions that cannot, in general, be absorbed into the standard hydrodynamic variables. These contributions lead to anisotropic effects in the fluid dynamics and can give rise to preferred-direction particle transport and additional forces and torques.

We also derived a first-order prescription to incorporate the Lorentz-violating corrections into a given Lorentz-invariant equation of state. This prescription expresses the modified thermodynamic variables in terms of their Lorentz-invariant counterparts, providing a practical way to evaluate the leading-order corrections without requiring a complete reformulation of the underlying equation of state.

The formalism developed here also opens several directions for future work. A complete $3+1$ decomposition and numerical implementation will be necessary to investigate the dynamical consequences of the derived equations in specific astrophysical settings. In particular, extending the present single-fluid description to a multicomponent plasma will be essential to establish the conditions under which conventional relativistic magnetohydrodynamic approximations remain applicable in the presence of Lorentz violation, and ultimately to develop a self-consistent Lorentz-violating RMHD framework. Further extensions could incorporate a photon fluid, allowing the investigation of coupled matter--radiation systems within a common relativistic description.

\begin{acknowledgments}

The authors are supported by the S\~{a}o Paulo Research Foundation (FAPESP) through grant number 2021/01089-1. VdS is supported by CNPq through grant number 308837/2023-1. MD is supported by CNPq through grant number 140114/2025-4. The authors acknowledge the National Laboratory for Scientific Computing (LNCC/MCTI,  Brazil) for providing HPC resources for the SDumont supercomputer (http://sdumont.lncc.br).

\end{acknowledgments}

\appendix

\section{Wigner function}
\label{sec:app1}

\subsection{Gauge-invariant form of the Wigner equation}

As shown in Eq.~\eqref{eq:W_EOM_1}, applying a Fourier transform to the object \(i\gamma \cdot (\tfrac{1}{2}\partial_x-\partial_y)\bar{\psi}U\psi\) and using the modified Dirac equation yields
\begin{equation}
\begin{aligned}
\left [m - \gamma\cdot (p + \tfrac{i\partial_x}{2}) \right]\hat{W}(x,p) =\\
ie \frac{\partial}{\partial p^\mu}\int \frac{d^4y}{(2\pi)^4}e^{-ip\cdot y} \bar \psi (x_+)P\left[\int_0^1 ds \left(1-s\right) F_{\mu \nu}(x_- -sy) U(A,x_+,x_-) \right]\gamma^\nu \psi(x_-) \ + \\
+ \frac{\mathcal{H}}{M_{\rm Pl}}\int \frac{d^4y}{(2\pi)^4}\bar\psi(x_+)PU(A,x_+,x_-)\left(n \cdot D_{x_-} \right)^2 \psi(x_-),
\end{aligned}
\end{equation}
The standard terms are already written in an explicitly gauge-invariant form, which is convenient for the subsequent c-number field approximation. The last term, however, contains the Lorentz-violating contribution and must be further manipulated to obtain a form suitable for the same procedure.

We first note that
\begin{equation}
    \begin{aligned}
       \bar{\psi}U\left(n\cdot D\right)^2\psi = \left(n\cdot \partial\right)^2\left[\bar\psi U \psi \right] + ie\bar\psi U \left( n\cdot\partial\right)\left(n\cdot A \right)\psi \\
       + 2ie\bar\psi U \left(n\cdot A \right)\left(n\cdot\partial \right)\psi - e^2\bar\psi U \left(n\cdot A \right)^2\psi\\
       - \bar\psi(n\cdot\partial)^2U\psi - 2\bar\psi\left(n\cdot\partial \right)U\left(n\cdot\partial \right)\psi,
    \end{aligned}
\end{equation}
where all derivatives act with respect to $x_-$. In order to perform the integrations by parts below, it is convenient to eliminate terms in which derivatives act exclusively on $\psi$. Rearranging the expression above, we obtain
\begin{equation}
    \begin{aligned}
       \bar{\psi}U\left(n\cdot D\right)^2\psi = \left(n\cdot \partial\right)^2\left[\bar\psi U \psi \right] + 2ie \left( n\cdot A\right) \left(n\cdot\partial \right)\left[\bar\psi U \psi \right] \\
       + ie\bar\psi U \left(n\cdot\partial\right)\left(n\cdot A\right)\psi -e^2\left(n\cdot A\right)^2\bar\psi U \psi \\
       +\bar\psi\left(n\cdot\partial\right)^2U \psi - 2ie\left(n\cdot A\right)\bar\psi \left(n\cdot\partial\right)U \psi \\
       -2\left(n\cdot\partial\right)\left[\bar\psi \left(n\cdot\partial\right)U\psi\right].
\end{aligned}
\end{equation}
The derivatives for $U(A,x_+,x_-)$ are given by
\begin{equation}
    \left(n\cdot\partial\right)U = ieU\times\left[\left(n\cdot A\right) - y^\nu \int_0^1 ds\left(1-s\right)n^\mu F_{\mu\nu} \right],
\end{equation}
\begin{equation}
\begin{aligned}
    \left(n\cdot\partial\right)^2U = ieU\times\Bigg\{ie\left[\left(n\cdot A\right) - y^\nu \int_0^1 ds\left(1-s\right)n^\mu F_{\mu\nu} \right]^2 \\
    + \left(n\cdot\partial\right)\left(n\cdot A\right) - y^\nu\int_0^1ds(1-s)^2n^\mu \left(n\cdot\partial\right)F_{\mu\nu} \\
    + \int_0^1ds(1-s)n^\mu n^\nu F_{\mu\nu}.
\end{aligned}   
\end{equation}

Using these relations and performing integrations by parts on the terms involving derivatives with respect to $y$, given that
$\partial_{x_-}=\tfrac{1}{2}\partial_x-\partial_y$, we obtain
\begin{equation}
\begin{aligned}
\left[m - \gamma \cdot (p + \tfrac{i\partial_x}{2})+ \frac{\mathcal{H}}{M_{\rm Pl}} \left(n \cdot p + \frac{i n\cdot \partial_x}{2} \right)^2 \right] \hat{W}(x,p) = \\
 ie \left[ \gamma^\nu \frac{\partial}{\partial p^\mu} - \frac{2\mathcal{H}}{M_{\rm Pl}} \left( n \cdot p + \frac{i n\cdot \partial_x}{2} \right) n^\nu \frac{\partial}{\partial p^\mu}\right] \\ \times \int \frac{d^4y}{(2\pi)^4}e^{-ip\cdot y} \bar \psi (x_+)P\left[\int_0^1 ds \left(1-s\right) F_{\mu \nu}(x_- -sy) U(A,x_+,x_-) \right]\gamma^\nu \psi(x_-) \ + \\
 -e\frac{\mathcal{H}}{M_{\rm Pl}}\frac{\partial}{\partial p^\mu}\int \frac{d^4y}{(2\pi)^4}e^{-ip\cdot y} \bar \psi (x_+) P \left[\int_0^1 ds \left(1-s\right)^2 \left(n \cdot \partial \right)n^\nu F_{\mu \nu}(x_- -sy) U(A,x_+,x_-) \right] \psi(x_-) \ + \\
 +e^2 \frac{\mathcal{H}}{M_{\rm Pl}} \left( \frac{\partial}{\partial p^\mu}\right)^2 \int \frac{d^4y}{(2\pi)^4}e^{-ip\cdot y} \bar \psi (x_+)P\left[\int_0^1 ds \left(1-s\right)n^\nu F_{\mu \nu}(x_- -sy) U(A,x_+,x_-) \right]^2 \psi(x_-)\Bigg\}.
\end{aligned}
\label{eq:app:wigner}
\end{equation}

This equation provides the gauge-invariant form of the Wigner equation used in Eq.~\eqref{eq:W_EOM_2}.

\subsection{Arbitrary c-number fields}

In this Appendix, we detail the algebraic steps leading to Eq.~\eqref{eq:W_EOM_3}. Following Refs.~\cite{VASAK1987462,Elze:1986hq,Elze:1986qd}, we replace the gauge field operator with its classical $c$-number ensemble average, $F(x) \equiv \langle F(x)\rangle$. Within the Wigner formalism, this $c$-number evaluation corresponds to the Hartree mean-field approximation, which neglects two-body correlations and quantum fluctuations. This treatment provides a mean-field description in which collective dynamics are captured through the classical electromagnetic field, while two-body correlations and quantum fluctuations are neglected. This approximation provides a minimal, analytically tractable framework for isolating leading-order LIV effects in the hydrodynamics of a relativistic charged fluid, establishing a foundation for future collisional or correlated extensions.

To evaluate the kinetic equations, we make use of the following operator identities:
\begin{equation}
    f(x+a) = e^{a\partial_x}f(x),
\end{equation}
\begin{equation}
    \int d^4y \ e^{-ip\cdot y}\ f(y)\ g(y) =  f(i\partial^p)\int d^4 e^{-ip\cdot y}\ g(y).
\end{equation}
Using these relations to replace the dependence on $y$ by derivatives with respect to $p$, and taking the field strength $F^{\mu\nu}(x)$ outside the integral, Eq.~\eqref{eq:app:wigner} becomes
\begin{equation}
    \begin{aligned}
        \left\{ m - \gamma\cdot\left(p + \tfrac{1}{2}i\partial_x\right) - e\left[ \tfrac{i}{2}j_0(\tfrac{\triangle}{2}) + \tfrac{1}{2}j_1(\tfrac{\triangle}{2})\right]\gamma^\nu F_{\mu\nu} \partial_p^\mu \right\}W(x,p) = \\
        = \frac{\mathcal{H}}{M_{\rm Pl}}\Big\{ -2e\left(n\cdot p + \tfrac{i}{2}n\cdot\partial_x\right)\left[ \tfrac{i}{2}j_0(\tfrac{\triangle}{2}) + \tfrac{1}{2}j_1(\tfrac{\triangle}{2})\right] n^\nu F_{\mu\nu} \partial_p^\mu \\
        -e\left(n\cdot\partial\right) \left[\tfrac{1}{3}j_0(\tfrac{\triangle}{2}) - \tfrac{i}{2}j_1(\tfrac{\triangle}{2}) - \tfrac{1}{6}j_2(\tfrac{\triangle}{2}) \right] n^\nu F_{\mu\nu} \partial_p^\mu\\
        +e^2\left[ \tfrac{i}{2}j_0(\tfrac{\triangle}{2}) + \tfrac{1}{2}j_1(\tfrac{\triangle}{2})\right]^2 \left(n^\nu F_{\mu\nu}\right)^2\left(\partial_p^\mu\right)^2 - \left(n\cdot p + \tfrac{i}{2}n\cdot\partial_x \right)^2\Big\} W(x,p),
    \end{aligned}
    \label{eq:app:cnumber}
\end{equation}
where $\triangle \equiv \partial^p \cdot \partial_x$ and $j_i(z)$ are the spherical Bessel functions.

Introducing the operators
\begin{equation}
\nabla^\mu = \partial_x^\mu - ej_0(\tfrac{\triangle}{2})F^{\mu \nu}\partial_\nu^p
\end{equation}
and
\begin{equation}
\Pi^\mu = p^\mu - \frac{1}{2}ej_1(\tfrac{\triangle}{2})F^{\mu \nu}\partial_\nu^p,
\end{equation}
Eq.~\eqref{eq:app:cnumber} can be recast in the compact form
\begin{equation}
\begin{aligned}
\left[\gamma \cdot \left( \Pi + \tfrac{i\nabla}{2} \right) - m \right]W(x,p) = \frac{\mathcal{H}}{M_{\rm Pl}}\left[n \cdot \left(\Pi + \tfrac{i\nabla}{2} \right) \right]^2 W(x,p) \ + \\
+\frac{\mathcal{H}}{M_{\rm Pl}}e\left\{ \frac{i}{6}j_0 (\tfrac{\triangle}{2}) + \frac{1}{2}j_1 (\tfrac{\triangle}{2}) - \frac{i}{3}j_2 (\tfrac{\triangle}{2}) \right\} \left[\tfrac{i n\cdot \partial_x}{2}; n_\mu F^{\mu \nu} \right] \partial_\nu^p \ W(x,p),
\end{aligned}
\end{equation}
where $[a;b]$ denotes the commutator between the operators $a$ and $b$.

This form provides the starting point for the Clifford decomposition of the Wigner function, which is used below to obtain the coupled equations for its scalar, vector, axial-vector, pseudoscalar, and tensor components.

\subsection{Clifford decomposition}

The Wigner function can be decomposed in the Clifford basis,
$\{{\mathbb{I},\gamma^5,\gamma^\mu,\gamma^\mu\gamma^5,\sigma^{\mu\nu}}\}$, as
\begin{equation}
W = \mathcal{F} + i\gamma^5\mathcal{P} + \gamma^\mu\mathcal{V}_\mu + \gamma^\mu \gamma^5\mathcal{A}_\mu + \frac{1}{2}\sigma^{\mu \nu}\mathcal{S}_{\mu \nu}.
\end{equation}
The Clifford components are obtained through the following projections:
\begin{equation}
    \mathcal{F} = \frac{1}{4}{\rm tr} \ W(x,p),
\end{equation}
\begin{equation}
    \mathcal{P} = -\frac{i}{4}{\rm tr} \ \gamma^5W(x,p),
\end{equation}
\begin{equation}
    \mathcal{V}_\mu = \frac{1}{4}{\rm tr} \ \gamma_\mu W(x,p),
\end{equation}
\begin{equation}
    \mathcal{A}_\mu = \frac{1}{4}{\rm tr} \ \gamma_5\gamma_\mu W(x,p),
\end{equation}
\begin{equation}
    \mathcal{S}_{\mu\nu} = \frac{1}{4}{\rm tr} \ \sigma_{\mu\nu}W(x,p).
\end{equation}

Defining
\begin{equation}
K^\mu \equiv \Pi^\mu + \frac{i}{2}\nabla^\mu,
\end{equation}
and applying the above projections, together with the following trace identities for the Dirac gamma matrices,
\begin{equation}
    \operatorname{tr}(\gamma^\mu)
    =
    \operatorname{tr}(\gamma^\mu\gamma^\nu\gamma^\alpha)
    =
    \operatorname{tr}(\gamma^5)
    =
    \operatorname{tr}(\gamma^5\gamma^\mu)
    =
    \operatorname{tr}(\gamma^5\gamma^\mu\gamma^\nu)
    =
    \operatorname{tr}(\gamma^5\gamma^\mu\gamma^\nu\gamma^\alpha)
    =
    0,
\end{equation}
\begin{equation}
    \operatorname{tr}(\mathbb{I}) = 4,
\end{equation}
\begin{equation}
    \operatorname{tr}(\gamma^\mu\gamma^\nu)
    =
    4g^{\mu\nu},
\end{equation}
\begin{equation}
    \operatorname{tr}(\gamma^\mu\gamma^\nu
    \gamma^\alpha\gamma^\beta)
    =
    4\left(
    g^{\mu\nu}g^{\alpha\beta}
    -
    g^{\mu\alpha}g^{\nu\beta}
    +
    g^{\mu\beta}g^{\nu\alpha}
    \right),
\end{equation}
\begin{equation}
    \operatorname{tr}(\gamma^5\gamma^\mu\gamma^\nu
    \gamma^\alpha\gamma^\beta)
    =
    -4i\epsilon^{\mu\nu\alpha\beta},
\end{equation}
we obtain the following system of equations for the Clifford components:
\begin{equation}
    \left( K \cdot \mathcal{V}\right) - m\mathcal{F} = \frac{1}{M_{\rm Pl}}\left\{ \left( n \cdot K\right)^2 + e\ j_B(\tfrac{\triangle}{2})\left[\tfrac{i n\cdot \partial_x}{2}; n_\mu F^{\mu \nu} \right]\partial^p_\nu \right\}\times\left\{\eta_1 \left(n \cdot \mathcal{V} \right) - \eta_2 \left(n\cdot\mathcal{A} \right) \right\},
\end{equation}
\begin{equation}
    i\left(K\cdot\mathcal{A} \right) + m\mathcal{P} = \frac{i}{M_{\rm Pl}}\left\{ \left( n \cdot K\right)^2 + e\ j_B(\tfrac{\triangle}{2})\left[\tfrac{i n\cdot \partial_x}{2}; n_\mu F^{\mu \nu} \right]\partial^p_\nu \right\}\times\left\{\eta_1 \left(n \cdot \mathcal{A} \right) - \eta_2 \left(n\cdot\mathcal{V} \right) \right\},
\end{equation}
\begin{equation}
\begin{aligned}
    K_\mu\mathcal{F} - ik^\nu\mathcal{S_{\mu \nu}} - m\mathcal{V}_\mu =
    \frac{1}{M_{\rm Pl}}&\left\{ \left( n \cdot K\right)^2 + e\ j_B(\tfrac{\triangle}{2})\left[\tfrac{i n\cdot \partial_x}{2}; n_\mu F^{\mu \nu} \right]\partial^p_\nu \right\} \\
    &\times \left\{\eta_1 n_\mu \mathcal{F} + \eta_2 n_\mu \mathcal{P} + 2n^\sigma\mathcal{S}_{\sigma \mu} + \frac{1}{2}n^\alpha \epsilon_{\mu \alpha \sigma \rho}\mathcal{S}^{\sigma \rho} \right\},
\end{aligned}
\end{equation}
\begin{equation}
\begin{aligned}
    iK_\mu \mathcal{P} + \frac{1}{2}\epsilon_{\mu\sigma \alpha \beta}K^\sigma \mathcal{S}^{\alpha\beta} - m\mathcal{A}_\mu = -\frac{1}{M_{\rm Pl}}&\left\{ \left( n \cdot K\right)^2 + e\ j_B(\tfrac{\triangle}{2})\left[\tfrac{i n\cdot \partial_x}{2}; n_\mu F^{\mu \nu} \right]\partial^p_\nu \right\} \\
    &\times\left\{\eta_1 \left(n \cdot \mathcal{V} \right) - \eta_2 \left(n_\mu\mathcal{F} + 2n^\sigma\mathcal{S_{\sigma \mu}} \right) \right\},
\end{aligned}
\end{equation}
\begin{equation}
\begin{aligned}
i\left(K_\nu\mathcal{V}_\mu -  K_\mu\mathcal{V}_\nu\right) + \epsilon_{\mu \nu\alpha\beta}K^\alpha\mathcal{V}^\beta - m\mathcal{S}_{\mu\nu} = \frac{1}{M_{\rm Pl}}\left\{ \left( n \cdot K\right)^2 + e\ j_B(\tfrac{\triangle}{2})\left[\tfrac{i n\cdot \partial_x}{2}; n_\mu F^{\mu \nu} \right]\partial^p_\nu \right\}\\
\times\left\{ \eta_1 \left(g_{\mu\nu}n\cdot\mathcal{V} + n_\nu\mathcal{V}_
\mu - n_\mu\mathcal{V}_
\nu\right) + n^\alpha \epsilon_{\mu\nu\alpha\sigma}\left(\eta_1\mathcal{A}^\sigma - \eta_2\mathcal{V}^\sigma\right)\right\},
\end{aligned}
\end{equation}
where $j_B(z) = \frac{i}{6}j_0 (z) + \frac{1}{2}j_1 (z) - \frac{i}{3}j_2 (z)$.

The above system can be separated into its real and imaginary parts, yielding two coupled sets of equations that must be solved simultaneously.

\section{Kinetic Derivation of the Energy-Momentum Conservation Law}
\label{sec:app:kinectic}

The system of equations governing the Clifford components can be employed to derive the kinetic transport equation.

Substituting Eq.~\eqref{eq:real:3} into Eq.~\eqref{eq:im:1}, we obtain
\begin{equation}
\begin{aligned}
    \frac{\nabla\cdot \Pi }{2m}\mathcal{F} + \frac{\nabla^\mu\nabla^\nu}{2m}S_{\mu\nu} -&\frac{D}{mM_{\rm Pl}}\nabla^\mu \left\{\eta_1n_\mu\mathcal{F} + \eta_2n_\mu\mathcal{P} + 2n^\sigma\mathcal{S}_{\sigma\mu} + n^\alpha \tilde{S}_{\mu\alpha} \right\} = \\
    &=\frac{\mathcal{Q}}{M_{\rm Pl}}\left\{\eta_1\left(n\cdot\mathcal{V}\right) -\eta_2\left(n\cdot\mathcal{A}\right)\right\}.
\end{aligned}
\end{equation}
Restoring the explicit dependence on $\hbar$, the above equation can be expressed as
\begin{equation}
    \begin{aligned}
        \hbar\left(\partial_x^\mu-eF^{\mu\nu}\partial_\nu^p\right)\left\{\left(p_\mu - \frac{\eta_1}{M_{\rm Pl}}\left(n\cdot p\right)^2n_\mu \right)\frac{\mathcal{F}}{2m}\right\} + \mathcal{O}(\hbar^2) = \\
        \hbar\frac{\eta_1}{M_{\rm Pl}}\left(\partial_x^\mu-eF^{\mu\nu}\partial_\nu^p\right)\left\{\left(n\cdot p\right)^2 n_\mu \frac{\mathcal{F}}{m} \right\} + \mathcal{O}(\hbar^2).
    \end{aligned}
\end{equation}
Dividing by $\hbar$ and retaining terms up to order $\hbar^0$, we obtain the LIV-modified transport equation
\begin{equation}
    \left(p_\mu - \frac{3\eta_1}{M_{\rm Pl}}\left(n\cdot p\right)^2n_\mu\right)\left(\partial_x^\mu-eF^{\mu\nu}\partial_\nu^p\right)\mathcal{F} = 0.
    \label{eq:app:transport}
\end{equation}

To establish the connection with the macroscopic conservation laws, we take the first momentum moment of Eq.~\eqref{eq:app:transport} by multiplying it by $p^\nu$ and adopting the c-number limit. This yields
\begin{equation}
    \int d^4p\left[p^\mu p^\nu - \frac{3\eta_1}{M_{\rm Pl}}\left(n\cdot p\right)^2n^\mu p^\nu\right]\partial_\mu^x\mathcal{F} = eF_{\mu}{}^\alpha\int d^4 p \left[p^\mu p^\nu - \frac{3\eta_1}{M_{\rm Pl}}\left(n\cdot p\right)^2n^\mu p^\nu\right]\partial_\alpha^p\mathcal{F}.
\end{equation}
Integrating the right-hand side by parts in momentum space, and assuming that the boundary terms vanish, we find
\begin{equation}
    \partial_\mu^x\int d^4p\left[p^\mu p^\nu - \frac{3\eta_1}{M_{\rm Pl}}\left(n\cdot p\right)^2n^\mu p^\nu\right]\mathcal{F} = -eF_{\mu}{}^\nu \int d^4 p \left[p^\mu- \frac{3\eta_1}{M_{\rm Pl}}\left(n\cdot p\right)^2n^\mu\right]\mathcal{F}.
\end{equation}
The left-hand side can be identified with the divergence of the fluid energy-momentum tensor defined in Eq.~\eqref{eq:integral:tensor}, while the right-hand side corresponds to the electromagnetic force density expressed in terms of the effective current defined in Eq.~\eqref{eq:integral:corrente}. Consequently, we recover the energy-momentum balance equation
\begin{equation}
    \partial_\mu T^{\mu\nu}_{\rm fluid} = j^{\mu}_{\rm eff}F^\nu{}_\mu,
\end{equation}
in agreement with the result previously obtained from Noether's theorem.

\section{Energy-momentum tensor properties}
\label{sec:app2}

\subsection{Electromagnetic sector}

The contribution of the electromagnetic sector to the canonical energy-momentum tensor is given by
\begin{equation}
    \begin{aligned}
        T^{\mu\nu}_{EM} = -F^{\mu\rho}&\partial^\nu A_\rho + \\
        &+\frac{\xi}{M_{\rm Pl}}\left[\frac{3}{2}n^\mu\left(n\cdot\partial\right) \left(n_\theta \tilde{F}^{\theta\rho}\right) - n^\rho\left(n\cdot\partial\right)\left(n_\theta \tilde{F}^{\theta\mu}\right)\right]\partial^\nu A_\rho + \\
        &+\frac{\xi}{M_{\rm Pl}}\left[ n^\sigma F_{\sigma\beta}\frac{n_\theta \epsilon^{\theta\beta\mu\rho}}{2}\left(n\cdot\partial\right) -  \left(n\cdot\partial\right)\left(n^\sigma F_{\sigma\beta}\right) \frac{n_\theta\epsilon^{\theta\beta\mu\rho}}{2} \right]\partial^\nu A_\rho + \\ 
        &+ \frac{\xi}{M_{\rm Pl}}\frac{n^\mu}{2}F_{\sigma\beta}\ n_\theta\ \partial^\nu\tilde{F}^{\theta\beta} - g^{\mu\nu}\mathcal{L}_\gamma.
    \end{aligned}
\end{equation}
Using the relation $\partial^\mu A_\nu = F^\mu{}_\nu + \partial_\nu A^\mu$, we rearrange the tensor so that the gauge-dependent terms are made explicit
\begin{equation}
    \begin{aligned}
        T^{\mu\nu}_{\rm EM} = -F^{\mu\rho}F^\nu{}_\rho - \partial_\rho\left(F^{\mu\rho}A^\nu \right) + \partial_\rho\left(F^{\mu\rho}\right)A^\nu \\
        +\frac{\xi}{M_{\rm Pl}}\Bigg\{ \Big[ \frac{3}{2}n^\mu\left(n\cdot\partial\right)n_\theta\tilde{F}^{\theta\rho} - n^{\rho}\left(n\cdot\partial\right)n_\theta \tilde{F}^{\theta \mu} \\
        + n^\sigma F_{\sigma\beta}n_\theta \frac{\epsilon^{\theta\beta\mu\rho}}{2}\left(n\cdot\partial\right) - \left(n\cdot\partial\right)n^\sigma F_{\sigma\beta}n_\theta \frac{\epsilon^{\theta\beta\mu\rho}}{2}\Big] F^\nu{}_\rho \\
        + \left[ \frac{3}{2}n^\mu\left(n\cdot\partial\right)n_\theta\tilde{F}^{\theta\rho} - n^{\rho}\left(n\cdot\partial\right)n_\theta \tilde{F}^{\theta \mu}\right]\partial_\rho A^\nu \\
        + \partial_\rho \left[\left( n^\sigma F_{\sigma\beta}n_\theta \frac{\epsilon^{\theta\beta\mu\rho}}{2}\left(n\cdot\partial\right) - \left(n\cdot\partial\right)n^\sigma F_{\sigma\beta}n_\theta \frac{\epsilon^{\theta\beta\mu\rho}}{2}\right)A^\nu\right] \\
        - \partial_\rho n^\sigma F_{\sigma\beta}n_\theta \frac{\epsilon^{\theta\beta\mu\rho}}{2}\left(n\cdot\partial\right) A^\nu + \partial_\rho \left(n\cdot\partial\right)n^\sigma F_{\sigma\beta}n_\theta \frac{\epsilon^{\theta\beta\mu\rho}}{2}A^\nu \\
        + \frac{n^\mu}{2}n^\sigma\partial^\nu\left(n_\theta \tilde{F}^{\theta\beta}\right) \Bigg\} - g^{\mu\nu}\mathcal{L}_\gamma.
    \end{aligned}
\end{equation}
Taking $\partial_\mu$ of the above equation, using the modified Maxwell equations, and applying the following relations:
\begin{equation}
    \partial_\mu F^{\nu}{}_\rho = \partial^\nu F_{\mu\rho} + \partial_\rho F^\nu{}_\mu,
\end{equation}
\begin{equation}
    \partial_\mu F_{\sigma\beta}\ \epsilon^{\theta\beta\mu\rho} = -\frac{1}{2}\partial_\sigma \tilde{F}^{\theta\rho},
\end{equation}
\begin{equation}
    \epsilon^{\theta\beta\mu\rho}\ \partial_\mu F^\nu{}_\rho = \frac{1}{2}\partial^\nu\tilde{F}^{\theta\beta},
\end{equation}
we obtain the divergence of the energy-momentum tensor of the electromagnetic sector:
\begin{equation}
   \begin{aligned}
       \partial_\mu T^{\mu\nu}_{\rm EM} = &-j^\rho_{\rm eff}F^{\nu}{}_\rho - j^\rho_{\rm eff}\partial_\rho A^\nu +\\
       &+\frac{\xi}{M_{\rm Pl}}\left[ \left(n\cdot\partial\right)\left(n_\theta\tilde{F}^{\theta\rho}\right);\partial^\nu\left(n^\sigma F_{\sigma\rho}\right)\right].
   \end{aligned}
   \label{eq:app:em_j}
\end{equation}
The explicitly gauge-dependent term, $-j^\rho_{\rm eff}\partial_\rho A^\nu$ is canceled by the corresponding fermionic contribution, as shown in the following subsection.

\subsection{Gauge invariance}

Upon calculating the canonical energy-momentum tensor for the fermionic sector, we obtain
\begin{equation}
    \begin{aligned}
        T^{\mu\nu}_{\rm f} =
        \frac{i}{2}\bar{\psi}\gamma^\mu\overleftrightarrow{\partial^\nu}\psi
        +
        \frac{n^\mu}{2M_{\rm Pl}}
        \bar{\psi}\mathcal{H}
        \left[
        \left(n\cdot \overleftrightarrow{D}\right)
        \overleftrightarrow{\partial^\nu}
        \right]
        \psi.
    \end{aligned}
\end{equation}
This tensor is not gauge invariant. We can rewrite all ordinary derivatives in terms of covariant derivatives, obtaining
\begin{equation}
    \begin{aligned}
        T^{\mu\nu}_{\rm f} &=
        \frac{i}{2}\bar{\psi}\gamma^\mu\overleftrightarrow{D^\nu}\psi
        +
        \frac{n^\mu}{2M_{\rm Pl}}
        \bar{\psi}\mathcal{H}
        \left[
        \left(n\cdot \overleftrightarrow{D}\right)
        \overleftrightarrow{D^\nu}
        \right]
        \psi
        \\
        &\qquad
        +e\bar\psi\gamma^\mu A^\nu \psi - \frac{ien^\mu}{M_{\rm Pl}}\bar\psi\mathcal{H}\left(n\cdot\overleftrightarrow{D}\right)A^\nu \psi.
    \end{aligned}
\end{equation}
The last two terms combine into the effective current, as defined in Sec.~\ref{sec:eom}. Therefore, taking the divergence of the fermionic energy-momentum tensor and using Eq.~\eqref{eq:conservacorrente}, we obtain
\begin{equation}
    \begin{aligned}
        \partial_\mu T^{\mu\nu}_{\rm f} &=
        \partial_\mu\left\{\frac{i}{2}\bar{\psi}\gamma^\mu\overleftrightarrow{D^\nu}\psi
        +
        \frac{n^\mu}{2M_{\rm Pl}}
        \bar{\psi}\mathcal{H}
        \left[
        \left(n\cdot \overleftrightarrow{D}\right)
        \overleftrightarrow{D^\nu}
        \right]
        \psi\right\}
        \\
        &\qquad
        +j_{\rm eff}^\mu \partial_\mu A^\nu.
    \end{aligned}
\end{equation}
The last term on the right-hand side is the opposite of the corresponding term in Eq.~\eqref{eq:app:em_j}. Therefore, the gauge-dependent contributions cancel between the electromagnetic and fermionic sectors, leaving a gauge-invariant expression for the divergence of the total energy-momentum tensor.

\section{Integral solutions}
\label{sec:app3}

\subsection{Modified Israel-Stewart variables}

As shown in Eq.~\eqref{eq:Israel}, the Israel-Stewart variables can be
written in terms of the energy as
\begin{equation}
I_{nq}
=
\frac{4\pi}{(2q+1)!!}
\int \frac{dp}{(2\pi)^3}\,
\frac{p^{2}}{E}\,
f(E)\,
E^{\,n-2q}\,
\left(\bar p^\mu\bar p_\mu\right)^q.
\end{equation}

In the following subsections, In the following subsections, where we perform the algebra required to obtain the modified energy-momentum tensor of the charged fluid, we
will also require the corresponding variables in which the
distribution function is replaced by its derivative with respect to
the energy. We define these modified Israel-Stewart variables as
\begin{equation}
I'_{nq}
=
\frac{4\pi}{(2q+1)!!}
\int  \frac{dp}{(2\pi)^3}\,
\frac{p^{2}}{E}\,
\frac{\partial f(E)}{\partial E}\,
E^{\,n-2q}
\left(\bar p^\mu\bar p_\mu\right)^q.
\end{equation}

Using $E\,dE=p\,dp$ and integrating by parts, we obtain
\begin{equation}
I'_{nq}
=
\frac{(-1)^q4\pi}{(2q+1)!!}
\int  \frac{dE}{(2\pi)^3}\, f(E)
\left[
(n-2q)E^{n-1-2q}p^{2q+1}
+
(2q+1)E^{n-2q+1}p^{2q-1}
\right].
\end{equation}

After rearranging the terms, this expression can be written as
\begin{equation}
I'_{nq}
=
\frac{(-1)^{q-1}4\pi}{(2q-1)!!}
\int  \frac{dE}{(2\pi)^3}\, f(E)
E^{n-2q+1}p^{2q-1}
-
\frac{(-1)^q4\pi}{(2q+1)!!}
\int  \frac{dE}{(2\pi)^3}\, f(E)
(n-2q)E^{n-2q-1}p^{2q+1}.
\end{equation}

Therefore, the modified variables can be expressed in terms of the
original Israel-Stewart variables as
\begin{equation}
I'_{nq}
=
I_{n-1,q-1}
-
(n-2q)I_{n-1,q}.
\label{eq:app:israeldf}
\end{equation}

\subsection{Ensemble relations}

In order to obtain the hydrodynamic limit, we require relations between ensemble
averages of gauge-invariant fermionic operators containing up to two
covariant derivatives and momentum integrals of the Wigner function.
These relations follow directly from the inverse Fourier transform of
the gauge-invariant Wigner operator,
\begin{equation}
\left\langle :
\bar{\psi}_{\beta}(x_+)
U(A,x_+,x_-)
\psi_{\alpha}(x_-)
: \right\rangle
=
\int d^4p \,
e^{ip\cdot y}
W_{\alpha\beta}(x,p).
\label{eq:app:wigner_inverse}
\end{equation}

For a local bilinear without derivatives, contracting
Eq.~\eqref{eq:app:wigner_inverse} with an arbitrary Dirac matrix
$\Gamma$ and taking the limit $y\to0$ gives
\begin{equation}
\left\langle
\bar{\psi}(x)\Gamma\psi(x)
\right\rangle
=
\operatorname{tr}
\int d^4p \,
\Gamma W(x,p),
\label{eq:app:bilinear_0deriv}
\end{equation}
where we have used $U(A,x_+,x_-)\to\mathbb{I}$ in the coincident-point
limit.

For operators containing one covariant derivative, acting with the
relative derivative $i\partial_\mu^y$ on
Eq.~\eqref{eq:app:wigner_inverse} generates a factor $-p_\mu$ on the
right-hand side. On the left-hand side, the derivative acting on the
fermion fields and the Wilson line gives, in the limit $y\to0$,
\begin{equation}
\lim_{y\to0}
\left[
i\partial^y_\mu\left(
\bar{\psi}_{\beta}(x_+)
U(A,x_+,x_-)
\psi_{\alpha}(x_-)
\right)\right]
=
-\frac{1}{2}
\bar{\psi}_{\beta}(x)
\left(i\overleftrightarrow{D_\mu}\right)
\psi_{\alpha}(x).
\end{equation}
Therefore,
\begin{equation}
\left\langle
\bar{\psi}(x)\Gamma
\left(i\overleftrightarrow{D_\mu}\right)
\psi(x)
\right\rangle
=
2\,\operatorname{tr}
\int d^4p \,
p_\mu \Gamma W(x,p).
\label{eq:app:bilinear_1deriv}
\end{equation}

Applying two successive relative derivatives in the same way gives
\begin{equation}
\begin{aligned}
\lim_{y\to0}
\left[
\left(i\frac{\partial}{\partial y^\mu}\right)
\left(i\frac{\partial}{\partial y^\nu}\right)
\bar{\psi}_{\beta}(x_+)
U(A,x_+,x_-)
\psi_{\alpha}(x_-)
\right]
=
\frac{1}{4}
\bar{\psi}_{\beta}(x)
\left(i\overleftrightarrow{D_\mu}\right)
\left(i\overleftrightarrow{D_\nu}\right)
\psi_{\alpha}(x),
\end{aligned}
\end{equation}
and
\begin{equation}
\left\langle
\bar{\psi}(x)\Gamma
\left(i\overleftrightarrow{D_\mu}\right)
\left(i\overleftrightarrow{D_\nu}\right)
\psi(x)
\right\rangle
=
4\,\operatorname{tr}
\int d^4p \,
p_\mu p_\nu \Gamma W(x,p).
\label{eq:app:bilinear_2deriv}
\end{equation}

These relations provide the mapping required to express the
ensemble-averaged fermionic energy-momentum tensor and effective
current in terms of momentum integrals of the Wigner function.

\subsection{Energy-momentum tensor}

Using the relations between ensemble averages and momentum-space integrals of the Wigner function, we can take the ensemble average of the fermionic energy-momentum tensor. Applying Eqs.~\eqref{eq:app:bilinear_1deriv} and \eqref{eq:app:bilinear_2deriv} to Eq.~\eqref{eq:fermionico}, together with Eq.~\eqref{eq:transport}, we obtain, to first order in the LIV parameter,
\begin{equation}
    \langle T^{\mu\nu}_{\rm f} \rangle
    =
    4\int d^4p
    \left[
        p^\mu p^\nu
        -
        \frac{3\eta_1}{M_{\rm Pl}}
        (n\cdot p)^2 n^\mu p^\nu
    \right]
    \frac{\mathcal{F}}{m}.
\end{equation}
Using the expression for the scalar component of the Wigner function in Eq.~\eqref{eq:Fmathcal}, we can perform the integration over $p^0$. In particular, expanding the Dirac delta function, we obtain
\begin{equation}
    \mathcal{F}
    =
    \frac{m}{(2\pi)^3}
    \frac{
        \theta(p^0)\delta(p^0-E_{\rm LIV})
    }{
        2p^0
        \left[
            1
            -
            \frac{3\eta_1}{M_{\rm Pl}}
            \frac{n^0}{p^0}
            (n\cdot p)^2
        \right]
    }
    f(x,p),
    \label{eq:app:dirac}
\end{equation}
where
\begin{equation}
    E_{\rm LIV}
    \simeq
    E
    +
    \frac{\eta_1}{M_{\rm Pl}}
    \frac{(n\cdot p)^3}{E}.
\end{equation}
The modified on-shell energy also affects the equilibrium distribution. Expanding the Fermi--Dirac distribution to first order in the LIV parameter gives
\begin{equation}
    f_{\rm FD}(E_{\rm LIV})
    \simeq
    f_{\rm FD}(E)
    +
    \frac{\eta_1}{M_{\rm Pl}}
    \frac{(n\cdot p)^3}{E}
    \frac{d f_{\rm FD}}{dE}.
    \label{eq:app:distr}
\end{equation}
Combining these contributions and performing the $p^0$ integration, we obtain the following covariant momentum-space expression for the modified energy-momentum tensor of the charged fluid:
\begin{equation}
    \begin{aligned}
        T^{\mu\nu}_{\rm fluid}
        =
        2
        \int
        \frac{d^3p}{(2\pi)^3(u\cdot p)}
        f(u\cdot p)
        \Bigg\{
        p^\mu p^\nu
        +
        \frac{\eta_1}{M_{\rm Pl}}
        (n\cdot p)^2
        \Bigg[
        \frac{3(u\cdot n)p^\mu p^\nu}{u\cdot p}
        -
        \frac{(n\cdot p)p^\mu p^\nu}{(u\cdot p)^2}
        \\
        +
        \frac{1}{f(u\cdot p)}
        \frac{\partial f(u\cdot p)}{\partial(u\cdot p)}
        \frac{(n\cdot p)p^\mu p^\nu}{u\cdot p}
        -
        3n^\mu p^\nu
        \Bigg]
        \Bigg\}.
    \end{aligned}
\end{equation}
From the available four-vectors, the energy-momentum tensor can be decomposed as
\begin{equation}
    T^{\mu\nu}_{\rm fluid}
    =
    A u^\mu u^\nu
    +
    B\Delta^{\mu\nu}
    +
    C\bar n^\mu\bar n^\nu
    +
    D_1 u^\mu\bar n^\nu
    +
    D_2 u^\nu\bar n^\mu.
\end{equation}
The coefficients can be determined by projecting $T^{\mu\nu}_{\rm fluid}$ onto the corresponding tensor structures, yielding
\begin{equation}
    A
    =
    T^{\mu\nu}_{\rm fluid}u_\mu u_\nu,
\end{equation}
\begin{equation}
    B
    =
    \frac{1}{2}
    T^{\mu\nu}_{\rm fluid}\Delta_{\mu\nu}
    -
    \frac{1}{2}
    \frac{
        T^{\mu\nu}_{\rm fluid}\bar n_\mu\bar n_\nu
    }{
        \bar n\cdot\bar n
    },
\end{equation}
\begin{equation}
    C
    =
    \frac{3}{2}
    \frac{
        T^{\mu\nu}_{\rm fluid}\bar n_\mu\bar n_\nu
    }{
        (\bar n\cdot\bar n)^2
    }
    -
    \frac{1}{2}
    \frac{
        T^{\mu\nu}_{\rm fluid}\Delta_{\mu\nu}
    }{
        \bar n\cdot\bar n
    },
\end{equation}
\begin{equation}
    D_1
    =
    \frac{
        T^{\mu\nu}_{\rm fluid}u_\mu\bar n_\nu
    }{
        \bar n\cdot\bar n
    },
\end{equation}
and
\begin{equation}
    D_2
    =
    \frac{
        T^{\mu\nu}_{\rm fluid}u_\nu\bar n_\mu
    }{
        \bar n\cdot\bar n
    }.
\end{equation}
Evaluating these projections in terms of the Israel--Stewart integrals, and using Eq.~\eqref{eq:app:israeldf} for the terms arising from the expansion of the equilibrium distribution, we obtain
\begin{equation}
    \begin{aligned}
        T^{\mu\nu}_{\rm fluid}
        =&
        \left(I_{20}-I_{21}\right)u^\mu u^\nu
        +
        I_{21}g^{\mu\nu}
        \\
        &+
        \frac{\eta_1}{M_{\rm Pl}}
        \Bigg\{
        \Big[
            (u\cdot n)^3 I_{3,-1}
            -
            4(u\cdot n)^3 I_{30}
            +
            3(u\cdot n)(\bar n\cdot\bar n)I_{30}
            -
            10(u\cdot n)(\bar n\cdot\bar n)I_{31}
        \Big]
        u^\mu u^\nu
        \\
        &\qquad
        +
        \Big[
            (u\cdot n)^3I_{30}
            +
            3(u\cdot n)(\bar n\cdot\bar n)I_{31}
        \Big]
        g^{\mu\nu}
        \\
        &\qquad
        +
        \Big[
            3(u\cdot n)^2I_{30}
            -
            3(\bar n\cdot\bar n)I_{32}
            -
            9(u\cdot n)^2I_{31}
            +
            3(\bar n\cdot\bar n)I_{31}
        \Big]
        u^\mu\bar n^\nu
        \\
        &\qquad
        -
        \Big[
            3(\bar n\cdot\bar n)I_{32}
            +
            9(u\cdot n)^2I_{31}
        \Big]
        u^\nu\bar n^\mu
        \Bigg\}.
    \end{aligned}
\end{equation}
Finally, we transform the energy-momentum tensor to the Landau frame, in which the fluid four-velocity $u^\mu_{\rm L}$ satisfies the eigenvalue equation
\begin{equation}
    T^{\mu\nu}
    \left(u_{\rm L}\right)_\nu
    =
    \epsilon_{\rm LIV}
    u^\mu_{\rm L}.
    \label{eq:app:landau}
\end{equation}
To first order in the LIV parameter, we write the Landau-frame velocity as
\begin{equation}
    u^\mu_{\rm L}
    =
    u^\mu
    +
    v^\mu,
\end{equation}
where $v^\mu$ represents the LIV correction. Substituting this expression into Eq.~\eqref{eq:app:landau} and keeping terms up to first order in $\eta_1/M_{\rm Pl}$, we find
\begin{equation}
    v^\mu
    =
    -
    \frac{\eta_1}{M_{\rm Pl}}
    \frac{
        3(\bar n\cdot\bar n)I_{32}
        +
        9(u\cdot n)^2I_{31}
    }{
        \epsilon_0+P_0
    }
    \bar n^\mu.
    \label{eq:app:v}
\end{equation}
Using Eq.~\eqref{eq:app:v} and expressing the tensor in terms of $u^\mu_{\rm L}$, while consistently retaining terms up to first order in the LIV parameter, we obtain
\begin{equation}
\begin{aligned}
T^{\mu\nu}_{\rm fluid}
=
&
(\epsilon_{\rm LIV}+P_{\rm LIV})
u^\mu_{\rm L}u^\nu_{\rm L}
-
P_{\rm LIV}g^{\mu\nu}
\\
&+
\frac{\eta_1}{M_{\rm Pl}}
\left[
    3(u_{\rm L}\!\cdot\!n)^2I_{30}
    +
    3(\bar n\!\cdot\!\bar n)I_{31}
\right]
u^\mu_{\rm L}\bar n^\nu,
\end{aligned}
\end{equation}
where
\begin{equation}
P_{\rm LIV}
=
P_0
-
\frac{\eta_1}{M_{\rm Pl}}
\left[
    (u_{\rm L}\!\cdot\!n)^3I_{30}
    +
    3(u_{\rm L}\!\cdot\!n)
    (\bar n\!\cdot\!\bar n)
    I_{31}
\right],
\end{equation}
and
\begin{equation}
\begin{aligned}
\epsilon_{\rm LIV}
=
&
\epsilon_0
+
\frac{\eta_1}{M_{\rm Pl}}
\Big[
    (u_{\rm L}\!\cdot\!n)^3I_{3,-1}
    -
    3(u_{\rm L}\!\cdot\!n)^3I_{30}
    \\
    &\qquad
    +
    3(u_{\rm L}\!\cdot\!n)
    (\bar n\!\cdot\!\bar n)
    I_{30}
    +
    7(u_{\rm L}\!\cdot\!n)
    (\bar n\!\cdot\!\bar n)
    I_{31}
\Big].
\end{aligned}
\end{equation}
Throughout the main text, and in particular in Eq.~\eqref{eq:fluidLIV}, we omit the subscript ``L'' and denote the Landau-frame four-velocity simply by $u^\mu$.

\subsection{Current}

The ensemble average of the effective current can be evaluated analogously. Using the relations in Eqs.~\eqref{eq:app:bilinear_0deriv} and \eqref{eq:app:bilinear_1deriv} for the current given in Eq.~\eqref{eq:current}, its semiclassical ensemble average can be written as
\begin{equation}
    \langle j^\mu_{\rm eff} \rangle
    =
    4e
    \int d^4p
    \left[
        p^\mu
        -
        3\frac{\eta_1}{M_{\rm Pl}}
        (n\cdot p)^2 n^\mu
    \right]
    \frac{\mathcal{F}}{m}.
\end{equation}
Using the expansion of $\mathcal{F}$ given in Eq.~\eqref{eq:app:dirac}, together with the correction to the equilibrium distribution in Eq.~\eqref{eq:app:distr}, we obtain the following momentum-space integral for the effective current including the LIV corrections:
\begin{equation}
    \begin{aligned}
        \langle j^\mu_{\rm eff} \rangle
        =
        4e
        \int
        \frac{d^3p}{2(u\cdot p)}
        f(u\cdot p)
        \Bigg\{
        p^\mu
        +
        \frac{\eta_1}{M_{\rm Pl}}
        (n\cdot p)^2
        \Bigg[
        \frac{3(u\cdot n)p^\mu}{u\cdot p}
        -
        \frac{(n\cdot p)p^\mu}{(u\cdot p)^2}
        \\
        +
        \frac{1}{f(u\cdot p)}
        \frac{\partial f(u\cdot p)}{\partial(u\cdot p)}
        \frac{(n\cdot p)p^\mu}{u\cdot p}
        -
        3n^\mu
        \Bigg]
        \Bigg\}.
    \end{aligned}
\end{equation}
For the effective current, the available four-vectors allow for the decomposition
\begin{equation}
    \langle j^\mu_{\rm eff} \rangle
    =
    Au^\mu
    +
    B\bar n^\mu.
\end{equation}
The corresponding coefficients are obtained by projecting the current onto the available vector structures, yielding
\begin{equation}
    A=\langle j^\mu_{\rm eff} \rangle u_\mu,
\end{equation}
and
\begin{equation}
    B
    =
    \frac{1}{\bar n\cdot\bar n}
        \langle j^\mu_{\rm eff} \rangle n_\mu
\end{equation}
Evaluating these coefficients in terms of the Israel--Stewart integrals, we obtain
\begin{equation}
    \begin{aligned}
        \langle j^\mu_{\rm eff}\rangle
        =
        &\,
        eI_{10}u^\mu
        \\
        &+
        \frac{e\eta_1}{M_{\rm Pl}}
        \Big[
            (u\cdot n)^3I_{2,-1}
            -
            6(u\cdot n)(\bar n\cdot\bar n)I_{21}
            +
            3(u\cdot n)(\bar n\cdot\bar n)I_{20}
            -
            (u\cdot n)^3I_{20}
        \Big]
        u^\mu.
    \end{aligned}
\end{equation}
Finally, applying the transformation to the Landau frame using the correction given in Eq.~\eqref{eq:app:v}, the current conservation equation takes the form
\begin{equation}
    \partial_\mu
    \left(
        \rho_{\rm LIV}u^\mu_{\rm L}
        +
        \mathcal{C}\bar n^\mu
    \right)
    =
    0,
\end{equation}
where
\begin{equation}
    \begin{aligned}
        \rho_{\rm LIV}
        =
        \rho_0
        +
        \frac{\eta_1}{M_{\rm Pl}}
        \Big[
        &
        (u_{\rm L}\cdot n)^3I_{2,-1}
        -
        4(u_{\rm L}\cdot n)^3I_{20}
        \\
        &
        -
        6(u_{\rm L}\cdot n)
        (\bar n\cdot\bar n)I_{21}
        +
        3(u_{\rm L}\cdot n)
        (\bar n\cdot\bar n)I_{20}
        \Big],
    \end{aligned}
\end{equation}
and
\begin{equation}
    \begin{aligned}
        \mathcal{C}
        =
        \frac{\eta_1}{M_{\rm Pl}}
        &
        \left(\frac{\rho_0}{P_0+\epsilon_0}\right)
        \left[
            3(\bar n\cdot\bar n)I_{32}
            +
            9(u_{\rm L}\cdot n)^2I_{31}
        \right].
    \end{aligned}
\end{equation}
As in the previous subsection, the subscript ``L'' is omitted throughout the main text, and the Landau-frame four-velocity is denoted simply by $u^\mu$.


\bibliographystyle{apsrev4-2}
\bibliography{apssamp}

\end{document}